\documentclass{jnmp}

\usepackage{amsmath}

\JNMPnumberwithin{equation}{section}

\newtheorem{theorem}{Theorem}[section]
\newtheorem{lemma}{Lemma}[section]
\newtheorem{proposition}{Proposition}[section]
\newtheorem{corollary}{Corollary}[section]

\theoremstyle{definition}

\begin{document}

%
\renewcommand{\evenhead}{G~Malykin, V~Pozdnyakova and I~Shereshevskii}
\renewcommand{\oddhead}{Random Groups in the Optical Waveguides Theory}

%
\thispagestyle{empty}

\FirstPageHead{8}{4}{2001}{\pageref{malykin-firstpage}--\pageref{malykin-lastpage}}{Article}

\copyrightnote{2001}{G~Malykin, V~Pozdnyakova and I~Shereshevskii}

\Name{Random Groups\\ in the Optical Waveguides Theory}
\label{malykin-firstpage}

\Author{G~B~MALYKIN~$^\dag$, V~I~POZDNYAKOVA~$^{\ddag*}$
and I~A~SHERESHEVSKII~$^{\ddag\star}$}

\Address{
$^\dag$~Institute of Applied Physics, Russian Academy of Sciences, \\
~~46 Uljanova street, Nizhny Novgorod,  RU-603600, Russia\\[10pt]
$^\ddag$~Institute for Physics of Microstructures,  Russian Academy of Sciences, \\
~~ GSP-105,  Nizhny Novgorod,  RU-603950, Russia\\
$^*$~E-mail: vera@ipm.sci-nnov.ru\\
$^\star$~E-mail: ilya@ipm.sci-nnov.ru}

\Date{Received February 9, 2001; Accepted May 10, 2001}

\begin{abstract}
\noindent
We propose a new approach to the mathematical description of light
propagation in a single-mode fiber light-guide (SMFLG) with random
inhomogeneities.  We investigate statistics of complex amplitudes of
the electric field of light wave by methods of the random group
theory.  We have analyzed the behavior of the coherence matrix of a
monochromatic light wave and the polarization degree of a
nonmonochromatic radiation in SMFLG with random inhomogeneities as the
length of the fiber tends to infinity; in particular, we prove that
limit polarization degree is equal to zero.
\end{abstract}

\section{Introduction}

To be able to analyze statistical parameters of light propagation in a
single-mode fiber light-guide (SMFLG) with random inhomogeneities is
vital in connection with broad application of SMFLG in promising
directions of industrial development such as optical communication
\cite{Okoshi, Dianov} and improvement of sensors of various physical
parameters \cite{18, 19, Lefevre, 8, 9, 10}.

Existence of random inhomogeneities in real optical fibers leads to
appearance of coupling of orthogonal polarization modes,  it is
accompanied by exchange of energy between modes.  In turn,  the mode
coupling causes random variations of a polarization degree of
radiation in SMFLG and,  as a result,  undesirable phenomena in devices
and communication lines developed on the base of SMFLG
\cite{Rashleigh,  Noda,  Poole1,  Gisin}.  For a rather long fiber even
small random inhomogeneities strongly affect radiation characteristics
and make their theoretical analysis quite complicated.

To investigate radiation characteristics in SMFLG with random
inhomogeneities physicists usually apply methods of perturbation
theory \cite{2,  3,  Kochar,  Burns83,  Shangyuan}.  These methods are,
however,  essentially restricted both in the magnitude of fluctuations
and the length of the fiber.  This forces one to interpretate the
results of such calculations for sufficiently long fibers.  As an
illustration we point at the discussion about the value (zero or not
zero?)  of the limit polarization degree in infinite fibers (see,
e.g.,  \cite{2,  3,  Kochar,  Burns83,  15,  model}).  It was this very
problem that initially prompted us to develop the methods used in this
paper. In particular, we have proved that the
limit polarization degree is zero (see Corollary \ref{Cor4} in Section 2).

Here we propose a new approach to the mathematical description and
investigation of light propagation in SMFLG with random
inhomogeneities.  Our methods provide with a correct analytical
investigation of statistical characteristics of complex amplitudes of
an electric field of light wave for arbitrarily long fibers.  Note,
nevertheless,  that the mathematical model (equation (\ref{eq})) we use
is only applicable for real physical devices with not too strong
fluctuations,  because it does not take into account the {\it
reflection} of light on inhomogeneities.  The results,  obtained in the
framework of our model,  present,  however,  the exact solution of {\it
mathematical} problems formulated.

Our model and methods give a convenient basis for
numerical simulation of different optical fiber devises (see,  e.g.,
\cite{model,  Intens,  p2_2,  p2_3,  p2_4}).

Three pillars our approach is based on are: the random process theory,
as an instrument for correct mathematical description of the
statistics of inhomogeneities in fiber, the differential equation with
random coefficients as mathematical model of wave propagation in
random media and random groups theory as (and it seems to be quite
unusual) an instrument for exact solving these equations.  Of similar
approaches to similar problems we only know~\cite{Tut}.

On the structure of the paper.  In Section~2 we introduce notions we need,
formulate the problems we intend to solve and our main results.  We
also formulate two theorems concerning the random groups theory
itself.  Both these theorems are needed to solve our problems; one of
them is only a reformulation in the form convenient to us of one of
the main results of the random groups theory \cite{Hannan, Grenander},
while the other one seems to be new, although very close to some known
results from the Markov process theory~\cite{Feller}.  Among
``physically meaningful" results we state that the limit polarization
degree is exactly zero.  All the proofs, both of physical and pure
mathematical nature, are collected in Section~3 divided into subsections.
Our calculations are sometimes quite complicated and sometimes simple
but rather long so we separated the assertions from their proofs.  We
included more or less detailed proofs because we believe that our
methods may be also of interest themselves.  In proofs we omit some
technical details when they do not contain in our opinion any
nontrivial idea.

\section{Basic definitions,  equations and main results}

We consider the propagation problem of a nonmonochromatic light wave
in a single-mode fiber light-guide (SMFLG) with varying directions of
anisotropic axes and without losses.

Let $z$ be the lengthwise coordinate in fiber,  $\beta$ a parameter
depending on wavelength and physical parameters of fiber,
$\displaystyle \vec{E}(\beta, z)={\binom{E_x(\beta, z)}{ E_y(\beta, z)}}$
the complex amplitude vector of the two orthogonal components of the
electric field of a monochromatic light wave at point $z$, where
$x$- and $y$-axes coincide with directions of anisotropic axes in  this point.  The
propagation of a monochromatic light wave in SMFLG is described by the
following equation for vector $\vec{E}(\beta, z)$ \cite{model, Monerie}:
\begin{equation} \label{eq}
\frac{\partial\vec{E}}{\partial z} = X_{\beta}\left(\Theta(z)\right)\vec{E},
\qquad
X_{\beta}\left(\Theta(z)\right) = \left( \begin{array}{cc}
\displaystyle \frac{{\rm i}\beta}{2} & \Theta(z) \vspace{2mm}\\
-\Theta(z) & \displaystyle -\frac{{\rm i}\beta}{2}
\end{array} \right),
\end{equation}
where $\Theta(z)$ is the rotation velocity (twist) of anisotropic
axes,  i.e,  the derivatives of the directions of anisotropic axes with
respect to $z$.

The real optical fibers have random inhomogeneities occasioned by
uncontrollable peculiarities in the preparation process of fiber and
fiber devices.  We will assume in what follows that the function
$\Theta(z)$ is a realization of a random process.  Observe that though
in real fibers there appears not only fluctuation of anisotropic
axes's directions,  we disregard all other kinds of random
inhomogeneities.  The reason is that their influence on light
propagation in fiber is not so essential as that of the considered
one,  see,  e.g.,  \cite{Monerie,  Payne,  Kaminov}.

Our main goal is to investigate the statistical characteristics of the
solutions of equation~(\ref{eq}).  To do so,  we need a rigorous
mathematics description of the process $\Theta(z)$.  It seems that the
statistical characteristics we are interested in do not depend
crucially on the details of this description.  Therefore,  we will
consider a sufficiently simple mathema\-tical model of such process,
slightly more general than suggested in~\cite{model}.  This model,  in
our mind,  is close enough to description of the real structure of
random inhomogeneities in fibers,  and,  on the other hand,  allows us to
obtain rigorous results. We give some additional arguments
in favour of our model in what follows.  Note also that this model
appears to be very useful in simulation of light propagation in random
SMLFG \cite{model,  Intens,  p2_2,  p2_3,  p2_4}.

Let $\{\Theta_k\}_{k=0}^\infty$ and
$\{l_k\}_{k=0}^\infty$ be two sequences of independent random
values such that all $\Theta_k$ are distributed with the same
probability measure $\mu_\Theta$ on $\mathbb{R}$ and $l_k$ are also
equivalently distributed with density $\rho_l$ supported on
$\mathbb{R}_+$.  Let the random function $\Theta(z)$ be defined by the
relation
\begin{equation} \label{model}
\Theta(z)=\Theta_k, \qquad {\rm where}\quad
\sum\limits_{j=1}^{k-1}l_j \leqslant z <
 \sum\limits_{j=1}^{k}l_j, \quad {\rm and}\quad k=1, 2, \dots\: .
\end{equation}
The random process (\ref{model}) is a simple and convenient
mathematical model,  its particular cases are often used in different
applications of the probability theory \cite{Rytov}.  As it was shown
in~\cite{model} this model with specific $\mu_\Theta$ and $\rho_l$ is
physically justified and results obtained in the frameworks of this
model are supported by empirical observations.

In what follows we denote by angular brackets $\langle\,\cdot\,\rangle $ the mean
value of a random value.

We will say that the {\it regular twist} of the fiber is {\it absent}
if the distribution of $\Theta$ is symmetric with respect to zero, in
particular, $\langle \Theta^{2n+1}\rangle=0$ for all non-negative integers~$n$.

Let us now introduce parameters of the light field in fiber,  the
parameters which are most interesting from a physical point of view.

{\it The coherence matrix} of a {\it monochromatic} light wave is
defined by the relation
\begin{equation} \label{J_mono}
J(\beta, z)=\vec{E}(\beta, z)\vec{E}^{\, \dag}(\beta, z),
\end{equation}
where $\dag$ denotes the Hermitean conjugation.  The mean coherence
matrix $\langle J(\beta, z)\rangle $ is an important characteristic of the
monochromatic wave,  because the coupling (energy exchange) of
polarization modes that occurs in SMFLG with random inhomogeneities
is described by the diagonal elements of this matrix.

In what follows we will need {\it the function $B(\beta)$ of
spectral density of nonmonochromatic radiation in fiber} by the
relation
\begin{equation}{\label{Spectr}}
B(\beta)=\mbox{tr}\, J(\beta, z).
\end{equation}
Note that due to equation (\ref{eq}) $B(\beta)$ is indeed independent
on~$z$.

A {\it incoherent nonmonochromatic} radiation in SMFLG is
characterized by its {\it polarization degree} $p(z)$ \cite{p},  which
is
\[
p(z) = \sqrt{1-\frac{4\det J(z)}{\mbox{tr}^2 J(z)}},
\]
where $J(z)$ is the {\it coherence matrix of a nonmonochromatic
radiation},
\begin{equation} \label{J}
J(z) = \int J(\beta, z)\, d\beta.
\end{equation}
The {\it mean square polarization degree} is then equal to
$\langle p^2(z)\rangle$.  The study of this value is one of the main goals of our
work.

Earlier,  in the frameworks of perturbation theory it was shown (see,
e.g., \cite{class_h}) that,  in the absence of the regular twist and if
$\vec{E}(\beta, 0)=\left(\begin{array}{c} 1\\ 0\end{array}\right)$,  the {\it averaged
intensities of the eigenmodes} in SMFLG as $z\to\infty$ are of the form
\begin{equation} \label{J_diag}
\langle J_{11}(\beta, z)\rangle =
\frac{1}{2}\left(1+\mbox{e}^{-2h(\beta)z}\right), \qquad
\langle J_{22}(\beta, z)\rangle =
\frac{1}{2}\left(1-\mbox{e}^{-2h(\beta)z}\right),
\end{equation}
where
\begin{equation} \label{h_class}
h(\beta) = \lim_{z\to\infty}\frac{1}{z}
\left\langle \left| \int_0^z\Theta(z)\mbox{e}^{\mbox{i}\beta
z}\, dz \right|^2\right\rangle
\end{equation}
is the {\it $h$-parameter} used usually to characterize the coupling
between polarization modes \cite{Rashleigh,  Kaminov, class_h,  Marcuse}.
As we will see in what follows,  the relation like
(\ref{J_diag}) is valid in our model,  but the formula for
$h$-parameter is slightly different from (\ref{h_class}) and coincides
with it only when the random twist is small enough.

We wish to analyze the behavior of these characteristics both as
$z\to\infty$ and at fini\-te~$z$.  It is easy to see that to calculate
$\langle J(\beta, z)\rangle $ and $\langle p^2(z)\rangle$ it is sufficient to know the
distribution of random vector $\vec{E}(\beta, z)$ and also joint
distribution of the vectors $\vec{E}(\beta_1, z)$ and $\vec{E}(\beta_2, z)$
at different $\beta_1$ and $\beta_2$.  We don't need other statistical
parameters of vectors $\vec{E}(\beta, z)$ and so we will investigate only
the mentioned ones.

Denote by $N(z)$ the nonnegative integer $N$ such
that
\begin{equation}\label{N(z)}
N(z)=\min\left\{N\mid \sum\limits_{j=1}^N l_j>z\right\}.
\end{equation}
Note that $N(z)$,  as a random integer function of $z$,  depends on the
realization of the random process $\Theta(z)$.  Now,  the solution of
equation (\ref{eq}) with function $\Theta(z)$ of the form (\ref{model})
can be expressed as
\begin{gather}
\vec{E}(\beta, z) = U(\beta, z)\vec{E}_0(\beta) \nonumber\\
\quad = M_{\beta}\left(
     z-\sum\limits_{k=1}^{N(z)-1}l_k, \Theta_{N(z)}\right)
     M_{\beta}\left(l_{N(z)-1}, \Theta_{N(z)-1}\right)\cdots
     M_{\beta}\left(l_1, \Theta_1\right) \vec{E}_0(\beta),\label{E(z)}
\end{gather}
where $\vec{E}_0(\beta)$ is a complex amplitude of light field at the
entry point of fiber,  and $M_{\beta}\left(l, \Theta\right)$ $=
\exp\left(l\,  X_{\beta}(\Theta)\right)$ is {\it Jones matrix}
\cite{Shurkliff} of the fiber section of length $l$ and fixed axis
twist $\Theta$,
\begin{equation} \label{Matr_M}
M_{\beta}(l, \Theta) = \left(\begin{array}{cc}
\displaystyle  \cos\left(\frac{l\beta_\theta}{2}\right) +
 \mbox{i}\frac{\beta}{\beta_\theta}\sin\left(\frac{l\beta_\theta}{2}\right)&
 \displaystyle \frac{2\Theta}{\beta_\theta}\sin\left(\frac{l\beta_\theta}{2}\right)
\vspace{2mm}\\
 \displaystyle-\frac{2\Theta}{\beta_\theta}\sin\left(\frac{l\beta_\theta}{2}\right)      &
\displaystyle \cos\left(\frac{l\beta_\theta}{2}\right) -
 \mbox{i}\frac{\beta}{\beta_\theta}\sin\left(\frac{l\beta_\theta}{2}\right)
                      \end{array} \right).
\end{equation}
Here we set
$\beta_\theta=\sqrt{\beta^2+4\Theta^2}$.

Random matrices $M_{\beta}(l, \Theta)$ generate a subgroup $G$ of the
group $SU(2)$ of $2\times 2$ unitary matrices.  Thus, the analysis of
the statistics of $\vec{E}(\beta, z)$ reduces to the analysis of
statistics of the products of random matrices from some group, i.e.,
to a problem from the theory of random groups.

In what follows we will need functions $m_{\beta k}(l, \Theta)$,
$k\in\{0, 1, 3\}$ such that
\begin{equation}\label{mDef}
   M_{\beta}(l, \Theta) =
  \left( \begin{array}{cc}
     m_{\beta 0}(l, \Theta) + \mbox{i}m_{\beta 1}(l, \Theta) &
     m_{\beta 3}(l, \Theta)\vspace{2mm}\\
    - m_{\beta 3}(l, \Theta)&
    m_{\beta 0}(l, \Theta) - \mbox{i}m_{\beta 1}(l, \Theta)
  \end{array} \right).
\end{equation}

Together with the random vector $\vec{E}(\beta, z)$ which describes the
complex amplitude of the electric field at point $z$,  we consider also
a random vector $\vec{E}_N(\beta)$ which describes the field at the
output of the fiber that consists of precisely $N$ random sections.
The vector $\vec{E}_N(\beta)$ is defined by a formula similar to
(\ref{E(z)}) but where the matrix-valued function $U(\beta, z)$ is
replaced with the matrix-valued function
\begin{equation} \label{U_N}
U_N(\beta) = M_{\beta}\left(l_N, \, \Theta_N\right)
                M_{\beta}\left(l_{N-1}, \, \Theta_{N-1}\right)
          \cdots M_{\beta}\left(l_1, \, \Theta_1\right).
\end{equation}
We will see that it is possible to use this ``discrete semigroup"
instead of ``continuous" one from (\ref{E(z)}) in order
 to investigate
the statistical properties of electric field in the random fiber.
This is useful because there exist powerful results about the
random matrix products in random group theory.

Let us formulate now some useful theorems based on the random group
theory.  Let $\{l_j\}_{j=1}^\infty$ be a sequence of
independent equivalently distributed on $\mathbb{R}_+$ random values
and $\{c_j\}_{j=1}^\infty$ be a sequence of independent
equivalently distributed random vectors with values in $\mathbb{R}^n$
and absolutely continuous with respect to Lebesgue measure probability
distributions.  Let also $Y:{\mathbb{R}}^n\longrightarrow
\mathfrak{g}$ be smooth enough and almost everywhere (with respect to
the measure $\mu_c$ generated by random vector $c$) nonconstant
function on ${\mathbb{R}}^n$ with values in the Lie algebra
$\mathfrak{g}$ of the orthogonal or unitary group acting in the
$m$-dimensional space, i.e., with values in the space of $m\times m$
skew-symmetric or skew-Hermitean matrices, respectively.  Define the
matrix-valued function $A:
{\mathbb{R}}_+{\times}{\mathbb{R}}^n\longrightarrow G$, where
$G=SO(m)$ or $G=SU(m)$, by setting
\[
A(z, q)=\exp\left(z\, Y(q)\right).
\]

The matrix-valued function $A(z, q)$ generates a {\it random semigroup},
i.e.,  the random matrix-valued function $U(z_1, z_2)$,  where $z_1, z_2\in
\mathbb{R_+}$,  of the form
\begin{gather*}
U(z_2, z_1) =A\left(z_2-\sum\limits_{j=1}^{N(z_2)-1} l_j, c_{N(z_2)}\right)\\
\qquad {}\times A\left(l_{N(z_2)-1}, c_{N(z_2)-1}\right)\cdots
A\left(l_{N(z_1)+1}, c_{N(z_1)+1}\right)
A\left(\sum\limits_{j=1}^{N(z_1)} l_j-z_1, c_{N(z_1)}\right) ,
\end{gather*}
where $N(z)$ is defined by (\ref{N(z)}) and $z_2\geqslant z_1\geqslant
0$.  (Hereafter we assume $l_0=0$.)  The term ``continuous random
semigroup" is related with the fact that a solution of the
stochastic differential equation (\ref{eq}) is of such form,  and also
with the obvious fact that the equality
$U(z_3, z_2)U(z_2, z_1)=U(z_3, z_1)$ holds for $z_3\geqslant
z_2\geqslant z_1\geqslant 0$.  Below we deal with the matrix-valued
function $U(z)\equiv U(z, 0)$.  We note that the defined above random
semigroup is stationary in the sense that the statistical
characteristics of the matrix $U(z_1, z_2)$ depend only on $z_2-z_1$.

Together with the {\it continuous} matrix semigroup,  we will consider
the {\it discrete} random semigroup $U_{NK}$, $N, K\in\mathbb{Z_+}$
which is defined for integer $N\geqslant K\geqslant 0$ by the equality
\[
U_{NK}= A(l_N, c_N)A(l_{N-1}, c_{N-1})\cdots A(l_{K+1}, c_{K+1})A(l_K, c_K) .
\]
As in the continuous case the term ``semigroup" stems from the
relation ${U_{NK}U_{KL}=U_{NL}}$ valid for $ N\geqslant K\geqslant
L\geqslant 0$,  the stationarity condition is realized and we will
again be interested in the statistical properties of the family
$U_N\equiv U_{N0}$.

We denote by $\widehat{G}$ the minimal closed subgroup in $ G$ which
contains all matrices
$\{A(t, q)\mid $ $(t, q)\in\mathbb{R}_+{\times}\,{\rm supp}\,\mu_c\}$.
By definition,  $\widehat{G}$ is
a non-discrete subgroup in the compact simple group $G$.

This theorem is a specialization of ``the central limit theorem" for
compact stochastic groups for the considered case (see,  for example,
\cite{Hannan, Grenander}).

\renewcommand{\footnoterule}{\vspace*{3pt}%
\noindent
\rule{.4\columnwidth}{0.4pt}\vspace*{6pt}}

\begin{theorem} \label{Th1} Assume that the family of matrices
$\left\{A(z, q)\mid(z, q)\in\mathbb{R}_+{\times}\, {\rm supp}\,\mu_c\right\}$
is not contained in any
conjugacy class with respect to any normal subgroup of $\widehat{G}$.  Then
the probability distribution of the random matrix $U_N$ tends to the
Haar's measure on $\widehat{G}$ as $N\to\infty$\footnote{ The
condition concerning the conjugacy classes is present in the central
limit theorem \cite{Hannan, Grenander} for arbitrary compact
stochastic group.  This condition is certainly realized if
$\widehat{G}$ is simple: no normal subgroups.
As we will see below,  if $\widehat{G}$ is semisimple this condition is
equivalent to the statement that the family
${\{A(z, q)\mid(z, q)\in\mathbb{R}_+{\times}\,{\rm supp}\,\mu_c\}}$ does not have any common
eigenvector corresponding to the eigenvalue different from $1$.}.
\end{theorem}

The following theorem includes several statements which make it possible to
calculate limit mean values of different linear (with respect to the
semigroups introduced) quantities.  Formally,  this theorem does not
depend on Theorem \ref{Th1} but in fact its hypothesis are
equivalent to the ones of Theorem \ref{Th1} and its conclusions are
corollaries of Theorem \ref{Th1}.

\begin{theorem}\label{Th2} Let $P_0$ be an orthogonal projection on a common
eigenspace corresponding to the common zero eigenvalue of the family
$\left\{Y(q)\mid q\in\,{\rm supp}\,\mu_c\right\}$.  Suppose that the characteristic function
$\hat{\rho}_l(\lambda)$, $\lambda\in\mathbb{C}$,  of measure $\rho_l$,
defined for $\mbox{\rm Re}\, {\lambda}\leqslant 0$ by the relation
\begin{equation}\label{hat_rho}
\hat{\rho}_l(\lambda)=\int_0^\infty\rho_l(z){\rm e}^{-\lambda z}\, dz=
\langle{{\rm e}^{-\lambda l}}\rangle,
\end{equation}
is holomorphic in the closed left half-plane and has a meromorphic
extension to the whole complex plane.  Then

1) the limit mean of matrices $U_N$ as $N\to\infty$
exists and
$
\lim\limits_{N\to\infty}\langle{U_N}\rangle =P_0$;

2) the limit mean of matrices $U(z)$ as $z\to\infty$
exists and $\lim\limits_{z\to\infty}\langle{U(z)}\rangle=P_0$.
\end{theorem}

This theorem allows us to find some limit values using discrete
semigroup $U_N$ instead of continuous $U(z)$.  As we see below,  this
is sometimes more simple to do.

The following statements are obtained by applying the limit theorems
on random matrix semigroups in order to derive statistical properties of
a radiation in SMFLG when the fiber length tends to infinity.

\begin{corollary}\label{Cor1}
The limit distribution of $\vec{E}_N(\beta)$ as
$N\to\infty$ is the uniform distribution on the three-dimensional
sphere.
\end{corollary}

\begin{corollary}\label{Cor2}
Let ${\beta_1\ne\pm\beta_2}$. Then the vectors $\vec{E}_N(\beta_1)$ and
$\vec{E}_N(\beta_2)$ are independent as $N\to\infty$.
\end{corollary}

\begin{corollary}\label{Cor3}
$\lim\limits_{z\to\infty}\langle{J(\beta, z)}\rangle=
\left(\begin{array}{cc} \displaystyle \frac{1}{2} & 0 \\ 0&\displaystyle \frac{1}{2}
\end{array}\right)$.
\end{corollary}

\begin{corollary}\label{Cor4}
$\lim\limits_{z\to\infty}\langle{p^2(z)}\rangle=0$.
\end{corollary}

These results deal with the limit properties of light waves in fibers.
The result of Corollary \ref{Cor3} is well known for fibers without
regular twist (see,  e.g.,  \cite{class_h}). Corollary~\ref{Cor4}
seems to be most meaningful for physicists.  Note in
this connection that the problem of determination of the limit
polarization degree is widely discussed in the literature up to now,
see,  e.g.,~\cite{2, 3,  Kochar,  Burns83,  15, model}.

It is very interesting also to find out the dependence of the mean
coherence matrix and polarization degree on the length-wise coordinate
$z$.  Some particular results in this direction are listed bellow.  We
emphasize that all these results have an asymptotic nature, i.e., are
valid approximately for {\it sufficiently large} $z$.

First,  some preliminary notices.

As mentioned above,  the problems connected with continuous semigroup
$U(z)$ are often essentially more complicated than the ones connected
with $U_N$.  This is why in what follows we deal with $U_N$ instead of
$U(z)$ and intend now to explain some reasons for possibility to
perform such replacement.  In fact,  we wish that there were no
essential differences between the asymptotic behavior of $\langle{U_N\xi}\rangle$
and $\langle U(N\langle l\rangle)\xi\rangle$ as $N\to\infty$.  We did not try to prove or
even to formulate this assertion in rigorous mathematical manner,  but
we base our wishes on various numerical experiments which confirmed
this ``fact'' \cite{Intens}.  We think that it is possible to make a
{\it theorem} from this ``intuitive" assertion,  but its proof must be
quite difficult.  The reader can anticipate this from presented in Section~3
proofs of asymptotic assertions for the discrete semigroup; they
appear to be long enough even in this relatively simple case.

\begin{proposition}\label{Prop1}
Suppose that in a fiber the
regular twist is absent and $2\langle m_{\beta 3}^2(l, \Theta)\rangle<1$,  where
$m_{\beta j}$ is defined in (\ref{mDef}).  Then the diagonal
elements of the ``mean coherence matrix"
$\left\langle U_N (\beta)J(\beta, 0)U_N^\dag(\beta)\right\rangle$ at large $N$ fulfill
relation (\ref{J_diag}) with $z=N\langle l\rangle$ and $h$-parameter of the
form
\begin{equation} \label{my_h}
  h(\beta)=-\frac{1}{2\langle l\rangle}\ln\left(1-
  \left\langle \frac{8\Theta^2}{\beta^2+4\Theta^2}
         \sin^2\left(\frac{l}{2}\, \sqrt{\beta^2+4\Theta^2}\right)\right\rangle
     \right).
\end{equation}
\end{proposition}

It is interesting to compare relation (\ref{my_h}) with
(\ref{h_class}).  To do so,  we must obtain an explicit formula for
the ``classical" $h$-parameter.  The following assertion holds.

\begin{proposition}\label{Prop2}
\begin{equation}\label{hClassEx}
\lim_{z\to\infty}\frac{1}{z}
\left\langle \left| \int_0^z\Theta(z)\mbox{e}^{\mbox{i}\beta z}\, dz \right|^2\right\rangle=
\frac{4\langle\Theta^2\rangle}{\beta^2\langle l\rangle}
\left\langle \sin^2\frac{l\beta}{2}\right\rangle.
\end{equation}
\end{proposition}

To compare the different expressions for $h$-parameter, i.e.,
(\ref{my_h}) and (\ref{hClassEx}), we consider the case when
$\rho_l(z)=\frac{1}{\langle l\rangle}\exp\left(-\frac{z}{\langle l\rangle}\right)$.
Suppose that ${\rm supp}\, \mu_\Theta \subset [-\Theta_{\max},
\Theta_{\max}]$ with ${\Theta_{\max}\ll\sqrt{\beta^2+\langle l\rangle^{-2}}}$.
Then it is easy to see that the ratio of expressions (\ref{my_h}) and
(\ref{hClassEx}) is of the form
$1+O\left(\frac{4\Theta_{\max}^2}{\langle l\rangle^{-2}+\beta^2}\right)$.  Thus,
the conditions of applicability of expression (\ref{h_class}) appear
to be quite rigidly restricted.

\begin{proposition}\label{Prop3} The
asymptotic of the mean square of the polarization degree is of the
form
\begin{equation} \label{asymptotics}
\langle p^2_N\rangle =\frac{1}{3} \sqrt{\frac{2\pi}{N}}\int
  \frac{\widetilde{B}^2(\beta)}{\sqrt{f(\beta)}}\, d\beta +
O\left(\frac{1}{N^{3/2}}\right),
\end{equation}
where $\widetilde{B}(\beta)=\frac{{\rm tr}\, J_0(\beta)}{{\rm tr}\,
J_N}=\frac{B(\beta)}{\int B(\beta)\, d\beta}$ is a normalized spectral
function,  $B(\beta)$ is defined in (\ref{Spectr}),  $f(\beta)$ is
a rational function of the averaged polynomials in $m_{\beta j}$ and
their derivatives with respect to $\beta$ (for details see proof in
Section~3).  In particular,  if the regular twist is absent, then
\begin{gather*}
f(\beta)=\frac{8}{3}\sum\limits_{k\in\{0, 1, 3\}}
\left\langle\left(\frac{\partial m_{\beta j}(l, \Theta)}{\partial \beta}\right)^2\right\rangle\\
\qquad {}+
  \frac{32}{3}\frac{\displaystyle \left\langle \frac{\partial m_{\beta 1}(l, \Theta)}{\partial \beta}
  m_{\beta 0}(l, \Theta)
-m_{\beta 1}(l, \Theta)
\frac{\partial m_{\beta 0}(l, \Theta)}{\partial
\beta}\right\rangle^2}{\langle m_{\beta3}^2(l, \Theta)\rangle}.
\end{gather*}
\end{proposition}

Observe, that beside the power series on $N$, the full asymptotic
expansion for $\langle {p^2_N}\rangle$ contains also a finite sum of terms of the
form $a{\rm e}^{-\alpha N}$.  (One can see it from the proof of this
Proposition.)  Unfortunately, we cannot explicitly estimate the values
$a$ and $\alpha$ in these terms using our approach, hence, we cannot
estimate the accuracy of asymptotic (\ref{asymptotics}) as function of~$N$.

\section{Proofs of basic assertions}

\subsection{Proof of Theorem \ref{Th2}}

To prove heading 1),  introduce matrix $S=\langle A\rangle $,  the mean
of the matrix family $\{A(z, q)\mid $ $
(z, q)\in\mathbb{R}_+{\times}\,{\rm supp}\, \mu_c\}$,  i.e.,
\[
S=\int \rho_l(z)
   \int \exp\left(z\, Y(q)\right)\, d \mu_c(q)\, dz.
\]
Since the families $\{l_j\}_{j=1}^\infty$ and
$\{c_j\}_{j=1}^\infty$ are independent,  we immediately see
that
\[
\langle U_N\rangle =S^N.
\]
Obviously,  $SP_0=P_0$,  therefore,  $S^NP_0=P_0$ and
$\lim\limits_{N\to\infty}S^NP_0=P_0$.

Let us show now that $\lim\limits_{N\to\infty}S^N(E-P_0)=0$,  wherefrom
the statement required.  To this end,  consider an arbitrary nonzero
vector $x\in \mathbb{C}^m$ such that $P_0x=0$. The
unitarity of $A$ implies that $P_0Sx=0$,  too.  Indeed,  if $Ax=x$,
then $x=A^{\dag}x$ and,  therefore,  $P_0=S^{\dag}P_0$.  Conjugating both
sides of this equality we see that $P_0=P_0S$.  It is clear that
absolute values of all eigenvalues of the matrix $S$ do not exceed $1$
since
\[
\|S\| = \|\langle A(z, q)\rangle \| \leqslant \langle \|A(z, q)\|\rangle  = 1.
\]
Now we will need a lemma.

\begin{lemma} \label{Lemma1}
(i) If the absolute value of the eigenvalue $\eta$ of
$S$ is equal to $1$,  then $\eta=1$;

(ii) the vector $\varepsilon$ is an eigenvector with eigenvalue
$1$ of $S$ if and only if it is a common eigenvector of the matrix
family $\left\{A(z, q)\mid  (z, q)\in\mathbb{R}_+{\times}\,{\rm supp}\,
\mu_c\right\}$ with eigenvalue $1$.
\end{lemma}

\begin{proof} Let $\eta\ne 1$ be the eigenvalue of the matrix $S$
and $\varepsilon_{\eta}$ be the corresponding norma\-li\-zed eigenvector.
Then there exists $q_0\in\,{\rm supp}\, \mu_c$ such that
\begin{equation} \label{Unconst}
\frac{d}{d z}\exp\left(z\, Y(q_0)\right)\varepsilon_{\eta}\ne 0,
\end{equation}
because otherwise for any $q\in\,{\rm supp}\, \mu_c$ we would have had
\[
0=\left\|\frac{d}{d
z}\exp\left(z\, Y(q)\right)\varepsilon_{\eta}\right\|=
\left\|\exp\left(z\, Y(q)\right)Y(q)\varepsilon_{\eta}\right\|=
\left\|Y(q)\varepsilon_{\eta}\right\| ,
\]
i.e.,  $Y(q)\varepsilon_{\eta}=0$ and $\eta=1$ in contradiction with
assumption.  By continuity of $Y$ the inequality (\ref{Unconst}) holds
in some neighborhood $\Omega_{q_0}$ of positive $\mu_c$-measure
of the point $q_0$.

There are two possibilities: either $\varepsilon_{\eta}$ is a common
eigenvector of the matrix family ${\left\{Y(q)\mid q\in\,{\rm supp}\, \mu_c\right\}}$ or
there exists a $q_0$ such that the vector
$\exp\left(z\, Y(q_0)\right)\varepsilon_{\eta}$ is non-collinear to
$\varepsilon_{\eta}$ for all nonzero $z$.

Consider the first possibility.  Since $Y$ is skew-Hermitian or
skew-ortho\-go\-nal, all its eigenvalues are imaginary.  Let ${\rm
i}\chi(q)$, where $\chi(q)\not\equiv 0$, be the eigenvalue of $Y(q)$
corresponding to the eigenvector $\varepsilon_{\eta}$.  From the
condition $\chi(q_0)\ne 0$ for some $q_0$ and a continuity of $Y$ it
follows that we can find a neighborhood $\Omega_{q_0}$
so that the inequality $|\chi(q)|>\frac{1}{2}|\chi(q_0)|$ holds for
any $q\in \Omega_{q_0}$.  Let $\hat{\rho}_l(\omega)$ be
a characteristic function of the distribution $\rho_l$, i.e.,
\[
    \hat{\rho}_l(\omega)=\int\rho_l(z)\, {\rm e}^{{\rm i}z \omega}\, d z.
\]
It is known that $|\hat{\rho}_l(\omega)|\leqslant 1 $ for all $\omega$ and
$|\hat{\rho}_l(\omega)|< 1 $ for  $\omega\ne 0$ if $\rho_l$ is piecewise
continuous function. So,  $|\hat{\rho}_l(\chi(q))|< 1-\delta $ for all
$q\in \Omega_{q_0}$ and some positive $\delta$.
Then
\begin{gather*}
|\eta|= \left|\left(S\varepsilon_{\eta}, \varepsilon_{\eta}\right)\right|
       = \left|\int d\mu_c(q)
\int{\rm e}^{{\rm i}z \chi(q)}\rho_l(z)\, d z \right|
       = \left|\int \hat{\rho}_l(\chi(q))\, d\mu_c(q) \right|
       \\
       \quad {}\leqslant  \int\left|\hat{\rho}_l(\chi(q))\right|\, d\mu_c(q)
       =  \left(\int_{\Omega_{q_0}}+
\int_{\mathbb{R}^n\backslash\Omega_{q_0}}\right)
|\hat{\rho}_l(\chi(q))|\, d\mu_c(q)
  \leqslant  1 - \delta\, \mu_c(\, \Omega_{q_0})< 1 ,
\end{gather*}
as was required.

The second possibility is considered in a similar way.  Let $z_0>0$ be
such that ${\rho_l(z_0)\ne 0}$.  Then,  by assumption,  the vector
$\exp\left(z_0\, Y(q_0)\right)\varepsilon_{\eta}$ is noncollinear to
$\varepsilon_{\eta}$ and for some ${\delta>0}$ we have
$\left|\left(\exp\left(z_0\, Y(q_0)\right)\varepsilon_{\eta}, \,
\varepsilon_{\eta}\right)\right| = 1-2\delta$.  It follows from the
continuity of $Y$ that
$\left|\left(\exp\left(z\, Y(q)\right)\varepsilon_{\eta}, \,
\varepsilon_{\eta}\right)\right|=1-\delta$ for all
$(z, q)\in[z_0-\Delta, \, z_0+\Delta]\times\Omega_{q_0}$.
Thus,
\begin{gather*}
|\eta|= \left|\left(S\varepsilon_{\eta}, \varepsilon_{\eta}\right)\right|
       = \left|\int d\mu_c(q)\int \rho_l(t)
\left(\exp\left(z\, Y(q)\right)\varepsilon_{\eta}, \,
\varepsilon_{\eta}\right) d z \right| \\
\qquad {}       \leqslant
1 - \delta\, \mu_c(\, \Omega_{q_0})\, \mu_l\left([z_0-\Delta, \,  z_0+\Delta]\right)
<  1.
\end{gather*}
This completes the proof of Lemma \ref{Lemma1}.
\end{proof}

These estimates and the theorem on the Jordan form of matrices
imply that there exists an $\nu<1$ such that the inequality
\[
\|Sx\|\leqslant \nu\|x\|
\]
holds for all $x$ such that $P_0x=0$.  Thus,  inequality
$\|S^Nx\|\leqslant \nu^N\|x\|$ and $S$-invariance of the projection
$P_0$ implies now that
\[
\lim\limits_{N\to\infty}S^N=P_0.
\]
This completes the proof of the first heading of Theorem~\ref{Th2}.

Let us prove heading 2).  The
independence of the random values $\{l_j\}_{j=1}^{\infty}$,
$\{c_j\}_{j=1}^{\infty}$ implies that
\begin{gather*}
\langle U(t)\rangle \equiv V(t)=\sum_{k=1}^{\infty} \int \prod\limits_{j=1}^{k} d
\mu_c(q_j)
\int_{\sum\limits_{j=1}^{k-1}l_j\leqslant z}
\prod\limits_{j=1}^{k-1}\rho_l(l_j)\, dl_j\\
\phantom{\langle U(t)\rangle \equiv} {}\times
\prod\limits_{j=1}^{k-1} {\rm e}^{l_j Y(q_j)}
{\rm e}^{\left(z-\sum\limits_{j=1}^{k-1}l_j\right)Y(q_k)}
\Phi_l\left(z- \sum\limits_{j=1}^{k-1}l_j\right),
\end{gather*}
where
\begin{equation}\label{Phi_l}
\Phi_l(z)=\int_{z}^\infty \rho_l(s)\, d s.
\end{equation}

Introduce the matrix
\[
    L(z)=\int {\rm e}^{z\, Y(q)}\, d \mu_c(q),
\]
and calculate the Laplace transform  $\widehat {V}(\lambda)$ of the
matrix-valued function $V(t)$,
\[
     \widehat{V}(\lambda)=\int_0^\infty V(z)\, {\rm e}^{-\lambda z}\, d
     z.
\]
To do this,  we use the convolution theorem for the Laplace transforms.
Thus,  setting
\[
\widehat{L}_\rho(\lambda)=\int_0^\infty \rho_l(z)L(z)\, {\rm
e}^{-\lambda z}\, d z,  \qquad
\widehat{L}_\Phi(\lambda)=\int_0^\infty \Phi_l(z)L(z)\, {\rm
e}^{-\lambda z}\, d z,
\]
we obtain the expression for $\widehat{V}(\lambda)$ of the form
\[
\widehat{V}(\lambda)=\sum_{k=1}^\infty\left(\widehat{L}_\rho(\lambda)\right)^{k-1
}\widehat{L}_\Phi(\lambda)
=\left(E-\widehat{L}_\rho(\lambda)\right)^{-1}\widehat{L}_\Phi(\lambda).
\]
Note that by the definition $\|\widehat{L}_\rho(\lambda)\|<1$ for
${\rm Re}\,\lambda >0$; hence,  all the singularities of the matrix-valued
function in last formula are only poles lying in left half-plane and
only $\lambda=0$ is a pole with the zero imaginary part.  (Recall
that as follows from our hypotheses about density $\rho_l$,  the
matrices $\widehat {L}_\rho(\lambda)$ and $\widehat{L}_\Phi(\lambda)$ are
meromorphic in the whole plane and regular in the closed right
half-plane.)

We now calculate the limit of $V(z)$ as $z\to\infty$ via the inverse
Laplace transformation.  For $a>0$ we have
\[
\lim\limits_{z\to\infty}V(z) = \lim\limits_{z\to\infty}
     \frac{1}{2\pi {\rm i}}\int_{a-{\rm i}\infty}^{a+{\rm
i}\infty}
     \widehat{V}(\lambda)\, {\rm e}^{\lambda z}\, d\lambda
= \mathop{\rm Res}\limits_{\lambda=0}\, \widehat{V}(\lambda).
\]
To calculate the residue in this formula,  observe that by hypotheses
the matrix $L(z)$ has a block structure,  $L(z)=P_0\oplus
L^\perp(z)$; hence,
$\widehat{L}_\rho(\lambda)=\hat{\rho}_l(\lambda)P_0\oplus
\widehat{L}^\perp_\rho(\lambda)$,  where $\hat{\rho}_l(\lambda)$ is defined
by (\ref{hat_rho}).  Thus,
\[
\widehat{V}(\lambda)=
   \left(1-\hat{\rho}_l(\lambda)\right)^{-1}P_0\widehat{L}_\Phi(\lambda)
   \oplus\widehat{V}^\perp(\lambda).
\]
Only the first term in this expression has a pole at $\lambda=0$,
and the corresponding residue is
\[
\mathop{\rm Res}\limits_{\lambda=0}\, \widehat {V}(\lambda)=\frac{P_0\widehat{L}_\Phi(0)}
{\displaystyle -\frac{d\hat{\rho}_l(\lambda)}{d\lambda}\Bigr|_{\lambda=0}}.
\]
It is evident that
\[
-\frac{d\hat{\rho}_l(\lambda)}{d\lambda}\Bigr|_{\lambda=0}=
\int_0^\infty z\rho_l(z)\, d z = \langle l\rangle,
\]
and since $P_0L(z)=P_0$,  we obtain
\[
P_0\widehat{L}_\Phi(0)=P_0\int_0^\infty\Phi_l(z)\, d z=\langle l\rangle P_0.
\]
Thus,   $\mathop{\rm Res}\limits_{\lambda=0}\,\widehat {V}(\lambda)=P_0$
which completes the proof.

\subsection{Proof of Corollary \ref{Cor1}}

The absolute value of vector $\vec{E}_N(\beta)$ does
not depend on $N$ due to (\ref{eq}).  Thus,  a distribution of the
vector $\vec{E}_N(\beta)$ has support on the three-dimensional sphere
of radius $|\vec{E}_0(\beta)|$.  The limit distribution of the
vector $\vec{E}_N(\beta)=U_N(\beta)\vec{E}_0(\beta)$ as $N\to\infty$ for
an arbitrary initial vector $\vec{E}_0(\beta)$,  is determined uniquely
by the limit distribution of the random matrix $U_N(\beta)$.

The random matrices $M_{\beta}(l, \Theta)$ defined by expression
(\ref{Matr_M}) are elements of the group $SU(2)$.  Let us show that
the subgroup ${\mathcal M}\subset SU(2)$ generated by the matrix family
${\left\{M_{\beta}(l, \Theta)\mid(l, \Theta)\in
\mathbb{R}_+{\times}\,{\rm supp}\,\mu_\Theta\right\}}$ coincides with $SU(2)$.
Then,  by Theorem~\ref{Th1},  the limit
distribution of $U_N(\beta)$ defined by (\ref{U_N}),  is the Haar
measure on $SU(2)$.  The matrices $M_{\beta}$ depend on two
independent random parameters: $l$ and $\Theta$.  Thus,
$\dim({\mathcal M})\geqslant 2$.
Since $\dim(SU(2))=3$,  either $\dim({\mathcal M})=2$ or
$\dim({\mathcal M})=3$.  In the last case ${\mathcal M}=SU(2)$.
It is known~\cite{Group},  that there is no two-dimensional subgroups in $SU(2)$.
We are done.

Thus,  the limit distribution of the random matrix $U_N(\beta)$ as
$N\to\infty$ is the Haar measure on $SU(2)$; hence,  the limit
distribution of the vector $\vec{E}_N(\beta) = U_N(\beta)\vec{E}_0(\beta)$
is uniform on the three-dimensional sphere with radius $|\vec{E}_0(\beta)|$.

\subsection{Proof of Corollary \ref{Cor2}}

We will show that the limit joint distribution of the
matrices $U_N(\beta_1)$ and $U_N(\beta_2)$ is the Haar measure on the
group ${SU(2)\times SU(2)}$.  This means that these matrices and
therefore the vectors $\vec{E}_N(\beta_1)$ and $\vec{E}_N(\beta_2)$ for
$\beta_1\ne\pm\beta_2$ became statistically independent as
${N\to\infty}$.

Let $G$ be a subgroup of ${SU(2)\times SU(2)}$.  The elements of $G$
are pairs of matrices $(g_1, g_2)$, ${g_{1, 2}\in SU(2)}$.  Denote by
${\rm Pr}_i: G\longrightarrow SU(2)$,  $i=1, 2$, the homomorphic projections,  i.e.,
${\rm Pr}_i\left((g_1, g_2)\right)=g_i$.  The following statement is known
from the theory of semisimple compact groups (see,  for example,~\cite{Group}).

\begin{lemma}\label{Lemma2}
Any automorphisms of the group $SU(2)$ is of the form
${\rm Aut}\,(g)=h f(g)h^{\dag}$,  where $h$ is some fixed element from $SU(2)$ and
$f(g)$ is equal to either $g$ or the complex conjugate matrix
$\overline{g}$.
\end{lemma}

As a corollary we obtain

\begin{lemma}\label{Lemma3}
 If the subgroup $G\subseteq SU(2)\times SU(2)$ is such
that both ${\rm Pr}_1$ and ${\rm Pr}_2$ are epimorphisms,  then either
$G=SU(2)\times SU(2)$ or $G=\left\{(g, {\rm Aut}\,(g))\mid   g\in SU(2)\right\}$,
where ${\rm Aut}$ is an automorphism of the group $SU(2)$.
\end{lemma}

\begin{proof}
Since ${\rm Pr}_{i}(G)=SU(2)$,  then $e\in{\rm Pr}_{i}(G)$ where $e$ is the unit
element of $SU(2)$.  Denote: $G_e={\rm Pr}_1^{-1}(e)$.  Then $G_e$ is a
normal subgroup in $SU(2)$ as the kernel of the homomorphism ${\rm Pr}_1$.

Set now ${G_e}^\prime=\left\{g\in SU(2)\mid  (e, g)\in G_e\right\}\subset
SU(2)$.  It is clearly that ${G_e}^\prime$ is a subgroup; let us show
that it is a normal one.  Indeed,  ${ghg^{-1}\in SU(2)}$ for any $h\in
{G_e}^\prime\subset SU(2)$ and given ${g\in SU(2)}$ because $SU(2)$ is
a group.  Since ${\rm Pr}_2$ is a surjection,  there exists $\tilde{g}$ such
that $(\tilde{g}, \ g)\in G$ and since $G_e$ is normal,  we see that
\[
{(\tilde{g}, \ g)(e, h)(\tilde{g}, \ g)^{-1}=\left(e,  ghg^{-1}\right)\in G_e}.
\]
Hence,  $ghg^{-1}\in {G_e}^\prime$ for any $g\in SU(2)$,  so
${G_e}^\prime$ is a normal subgroup.

But $SU(2)$ is a simple group,  so either ${G_e}^\prime=SU(2)$ or
${G_e}^\prime=e$.

Consider the first case.  Let $g_1\in SU(2)$.  Then since ${\rm Pr}_2$ is
onto,  there exists ${g^\prime\in SU(2)}$ such that $(g_1, g^\prime)\in
G$.  Let now $g_2\in SU(2)$; then $\left(e, g_2(g^\prime)^{-1}\right)\in
G_e\subset G$,  since ${g_2(g^\prime)^{-1}\in SU(2)={G_e}^\prime}$.
Therefore
\[
\left(e, g_2(g^\prime)^{-1}\right)\cdot(g_1, g^\prime)=(g_1, g_2)\in G.
\]
This means that $G=SU(2) \times SU(2)$.

Now,  consider the case when ${G_e}^\prime=e$.  In this case for
each $g\in SU(2)$ there is only one $g^\prime\in SU(2)$ such that
$(g, g^\prime)\in G$.  Indeed,  let there exist $g^{\prime\prime}\in
SU(2)$ such that $g^\prime\ne g^{\prime\prime}$ and
$(g, g^{\prime\prime})\in G$.  Then
$(g, g^\prime)^{-1}\cdot(g, g^{\prime\prime})\in G$ since $G$ is a
group.  On the other hand
\[
(g, g^\prime)^{-1}\cdot(g, g^{\prime\prime})=
\left(g_1^{-1}, (g^\prime)^{-1}\right)\cdot(g_1, g^{\prime\prime})=
\left(e, (g^\prime)^{-1}g^{\prime\prime}\right)\in G_e.
\]
Therefore,  $(g^\prime)^{-1}g^{\prime\prime}\in {G_e}^\prime=e$ and
$g^\prime=g^{\prime\prime}$.

Hence, there exists a map $f: SU(2)\longrightarrow SU(2)$ such that
$f(g)=g^\prime$ if and only if ${(g, g^\prime)\in G}$.  The map $f$ is
a homomorphism. Indeed, let $(g,g')\in G$ and $(h,h')\in G$.  Then
$(gh,g'h')\in G$ because of $G$ is a group and by the definition of
$f$ we see that $f(gh)=g'h'=f(g)f(h)$.

Let us show that $f$ is an injective homomorphism, i.e., if $g_1\ne
g_2$, then $f(g_1)\ne f(g_2)$.  The kernel of $f$ is a normal subgroup
of $SU(2)$.  This means that either ${\rm ker}\, f=e$ and in this case $f$ is
an injective homomorphism, or ${\rm ker}\, f=SU(2)$.  In the last case the
image of $SU(2)$ is the single element $e$ which is impossible because
${\rm Pr}_2(G)=SU(2)$.  Since ${\rm Pr}_2$ is onto, $f$ is a
surjective homomorphism.  So $f$ is both an injective and surjective,
hence, $f$ is an automorphism and ${G=\left\{(g, f(g))\mid g\in
SU(2)\right\}}$.  Proof of Lemma~\ref{Lemma3} is completed.
\end{proof}

Let now $G\subseteq SU(2)\times SU(2)$ be the group generated by
pairs of matrices $(M_{\beta_1}(l, \Theta)$, $M_{\beta_2}(l,
\Theta))$.  Thus, by Theorem \ref{Th1}, the limit distribution of
the matrix pair $(U_N(\beta_1)$, $U_N(\beta_2))$ as
$N\to\infty$ is the Haar measure on $G$.  Let us show that in our case
$G=SU(2)\times SU(2)$.  Since random matrices $M_{\beta}(l, \Theta)$
generated the whole $SU(2)$, then $G$ satisfies to the hypothesis of
Lemma~\ref{Lemma3}.  Assume that $G$ is different from $SU(2)\times
SU(2)$.  This means that there is an automorphism $\Psi$ of the group
$SU(2)$ such that $G=\left\{(g, \Psi(g))\mid g\in SU(2)\right\}$.  In
this case the Haar measure on $G$ is the image of the Haar measure on
$SU(2)$ with respect to the natural isomorphism $g\rightarrow(g,
\Psi(g))$ of groups $SU(2)$ and $G$.  In accordance with Lemma~\ref{Lemma2},
there exists a matrix $h\in SU(2)$ such that $g_2=h
f(g_1)h^{\dag}$ for any pair $(g_1, g_2)\in G$.

Assume that $f(g)=g$. For a given complex matrix $R$
we have
\[
\left\langle g_1Rg_2^{\dag}\right\rangle=
\left\langle g_1Rhg_1^{\dag}h^{\dag}\right\rangle=
\left(\int_{SU(2)} gRhg^{\dag}\, dg\right)h^{\dag},
\]
where $\langle \,\cdot\,\rangle $ denotes the averaging over Haar measure on $SU(2)$.
Any element $g\in SU(2)$ may be uniquely represented in the form
\begin{equation} \label{G}
g=\left(\begin{array}{rr}
   a & b \vspace{1mm}\\ -\bar{\mathstrut b}&\bar{\mathstrut a}
\end{array}\right),
\end{equation}
where $a$ and $b$ are complex numbers such that $|a|^2+|b|^2=1$.
Introduce the angles $\varphi$, $\psi$ and $\vartheta$ from the relations
\[
{a={\rm e}^{{\rm i}\varphi}\cos\vartheta}, \qquad
  {b={\rm e}^{{\rm i}\psi}\sin\vartheta},  \qquad
  {\vartheta\in[0, \pi/2]}, \qquad {\varphi, \psi\in[0, 2\pi]};
\]
and denote the elements of the matrix $R h$ by $r_{mn}$. Then
\[
gRhg^{\dag}=
\left(\begin{array}{rr}
   a\bar{{\vphantom b}a}r_{11}+a\bar{b}r_{12}+\bar{{\vphantom b}a}br_{21}+b\bar{b}r_{22} &
  -a br_{11}+a^2r_{12}-b^2r_{21}+a br_{22}\vspace{1mm}\\
  -\bar{{\vphantom b}a}\bar{b}r_{11}-\bar{b}^2r_{12}+\bar{{\vphantom b}a}^2r_{21}+
   \bar{{\vphantom b}a}\bar{b}r_{22}&
   b\bar{b}r_{11}-a\bar{b}r_{12}-\bar{{\vphantom b}a}br_{21}+a\bar{{\vphantom b}a}r_{22}
\end{array}\right).
\]
We need now to calculate
\[
\left\langle a\vphantom{b}^{k_1}\bar{\vphantom{b}a}
     \vphantom{b}^{k_2} b^{k_3}\bar b^{k_4}\right\rangle =
\left\langle (\cos\vartheta)^{k_1+k_2}(\sin\vartheta)^{k_3+k_4}
    {\rm e}^{{\rm i}(k_1-k_2)\varphi}{\rm e}^{{\rm i}(k_3-k_4)\psi}\right\rangle.
\]
The invariant normalized measure on $SU(2)$ is
$d g=\frac{1}{4\pi^2}\sin2\vartheta\, d\vartheta\, d\varphi\, d\psi$,
therefore,
\begin{equation} \label{<ab>}
\left\langle a\vphantom{b}^{k_1}\bar{\vphantom{b}a}
    \vphantom{b}^{k_2} b^{k_3}\bar b^{k_4}\right\rangle=\left\{
\begin{array}{lccc}
0                              & \mbox{if} \ \ k_1\ne k_2 &\mbox{or}&
k_3\ne k_4,\\
\displaystyle \frac{k_1!\, k_3!}{(k_1+k_3+1)!}& \mbox{if} \ \ k_1 = k_2  &\mbox{and}
& k_3 = k_4.
\end{array}\right.
\end{equation}
Using (\ref{<ab>}) we see that $\left\langle gRhg^{\dag}\right\rangle=\frac{1}{2}\,{\rm tr}\,(R
h)E$,  where $E$ is the unit matrix.  Therefore,  for $R=h^{-1}$ we have
\begin{equation}\label{First}
   \left\langle g_1h^{-1}g_2^{\dag}\right\rangle=h^{\dag}.
\end{equation}

Now we find the mean value of the product $g_1Rg_2^{\dag}$ in a
different way.  In the space $\mbox{Mat}(2; \mathbb{C})$ of complex
$2{\times}2$-matrices,  consider a linear operator $V$ of the form
\begin{equation} \label{V(r)}
V(R)=M^{\vphantom{\dag}}_{\beta_1}(l, \Theta)RM^{\dag}_{\beta_2}(l, \Theta).
\end{equation}
Then
$\left\langle U^{\vphantom{\dag}}_N(\beta_1)RU_N^{\dag}(\beta_2)\right\rangle=
\langle V\rangle^N(R)$
and $\left\langle g_1Rg_2^{\dag}\right\rangle =\lim\limits_{N\to\infty}\langle V\rangle^N(R)$.  It is
easy to see that $V$ preserves the inner product in the space of
complex matrices given by the formula ${(R_1,
R_2)={\rm tr}\,(R_1R_2^{\dag})}$.  Therefore, $V$ is a unitary operator.
Consider the matrix-valued function
${R(l)=M^{\vphantom{\dag}}_{\beta_1}(l, \Theta)RM^{\dag}_{\beta_2}(l,
\Theta)}$.  Since $M_{\beta}(l, \Theta)=\exp\left(l\,
X_{\beta}(\Theta)\right)$, function $R(l)$ satisfies the differential
equation
\[
\frac{d R(l)}{d l} = X^{\vphantom{\dag}}_{\beta_1}(\Theta)R(l)
                     + R(l)X^{\dag}_{\beta_2}(\Theta) = Y(\Theta)R(l)
\]
with the initial condition $R(0)=R$. This means that operator
(\ref{V(r)}) is of the form ${V=\exp\left(l\, Y(\Theta)\right)}$.

The matrix $\langle V\rangle$ satisfies the hypothesis of Theorem~\ref{Th2},  so
the limit of the expression $\langle V\rangle ^N(R)$ as $N\to\infty$ exists and is
equal to $P_0(R)$,  where $P_0$ is the projection on the common kernel
of the operators $Y(\Theta)$,  or equivalently,  the common eigenspace
of the operator family (\ref{V(r)}),  corresponding to the eigenvalue~1.
Let us show that the operator family (\ref{V(r)}) does not have the
common eigenvalue 1 when $\beta_1\ne\pm \beta_2$.  It is obvious that
if a matrix $R$ is an eigenvector of operator (\ref{V(r)}) with
eigenvalue 1 for any pair $(l, \Theta)\in
\mathbb{R}_+{\times}\,{\rm supp}\, \mu_\Theta$,  then the identity
\begin{equation} \label{R_eig}
M_{\beta_1}(l, \Theta)RM^{\dag}_{\beta_2}(l, \Theta)=R
\end{equation}
holds.  Differentiating both parts of (\ref{R_eig}) with respect to
$l$ and using the relation
$M_{\beta}(l, \Theta)$ $=\exp\left(l\, X_{\beta}(\Theta)\right)$,  we see
that
\[
X_{\beta_1}(\Theta)R + RX^{\dag}_{\beta_2}(\Theta) =0.
\]
The matrix $X_{\beta}(\Theta)$ defined by formula (\ref{eq}) can be
expressed as
\[
X_{\beta}(\Theta) = \frac{{\rm i}\beta}{2}\, \sigma_1 +
\Theta\, \sigma_2,
\]
where
$
\sigma_1=\left(\begin{array}{rr} 1&0\\0&-1 \end{array}\right)$ and
$
\sigma_2=\left(\begin{array}{rr} 0&1\\-1&0 \end{array}\right)$. Then
\begin{equation} \label{dR_dl}
  \left(\frac{{\rm i}\beta_1}{2}\, \sigma_1 + \Theta\, \sigma_2\right)R -
R\left(\frac{{\rm i}\beta_2}{2}\, \sigma_1 + \Theta\, \sigma_2\right)  = 0.
\end{equation}
Differentiating (\ref{dR_dl}) by $\Theta$ we obtain
\begin{equation} \label{R_s2}
\sigma_2 R - R\sigma_2 = 0.
\end{equation}
If we substitute (\ref{R_s2}) into (\ref{dR_dl}) we see that
\begin{equation} \label{R_s1}
\beta_1\sigma_1 R - \beta_2 R\sigma_1 = 0.
\end{equation}
It follows from relations (\ref{R_s2}) and (\ref{R_s1}) that the
elements $r_{ij}$ of the matrix $R$ satisfy the following system of
equations
\[
\left\{\begin{array}{l}
r_{11}-r_{22} = 0,\\
r_{12}+r_{21} = 0,\\
(\beta_1-\beta_2)r_{11} = 0,\\
(\beta_1+\beta_2)r_{12} = 0.
\end{array}\right.
\]
It is obvious that this system has a nontrivial solution only if
$\beta_1=\pm\beta_2$ and, therefore, $1$ is not a common eigenvalue of
the operator family (\ref{V(r)}) when $\beta_1\ne\pm\beta_2$.

Thus, it follows from heading 1) of Theorem \ref{Th2} that
\[
\left\langle g_1Rg_2^{\dag}\right\rangle=\lim_{N\to\infty}
\left\langle U_N(\beta_1)RU_N^{\dag}(\beta_2)\right\rangle=0
\]
for $\beta_1\neq\pm\beta_2$ and any complex matrix $R$ in
contradiction with~(\ref{First}).

The case when the automorphism  defining $G$  contains complex
conjugation can be considered in similar way.

So the joint distribution of matrices $U_{\infty}(\beta_1)$ and
$U_{\infty}(\beta_2)$ is the Haar measure on ${SU(2)\times SU(2)}$,
i.e.,  matrices $U_{\infty}(\beta_1)$ and $U_{\infty}(\beta_2)$
as well as vectors $\vec{E}_{\infty}(\beta_1)$,
$\vec{E}_{\infty}(\beta_2)$ are independent for different $\beta_1\ne
\pm \beta_2$.

\subsection{Proofs of Corollaries \ref{Cor3} and \ref{Cor4}}

Observe that,  due to equation (\ref{eq}),  {\it the total energy
of radiation},  being equal to $\int B(\beta)\, d\beta$,  where $B(\beta)$
is defined by (\ref{Spectr}),  is preserved during
propagation of the radiation in SMFLG,  so we may assume that $\int
B(\beta)\, d\beta = 1$.

Taking into account that $\vec{E}_N(\beta)=U_N(\beta)\vec{E}_0(\beta)$ we
rewrite expression (\ref{J}) for the coherence matrix at the output of
the fiber that consists of $N$ random sections as
\begin{equation} \label{J_N}
           J_N = \int
           U_N^{\phantom{\dag}}(\beta)\vec{E}_0^{\phantom{\dag}}(\beta)
           \vec{E}_0^{\, \dag}(\beta)U_N^{\dag}(\beta)\, d\beta.
\end{equation}
Since the trace of the product of two matrices does not depend on the
order of factors,  and the matrix $U_N(\beta)$ is unitary,  we
deduce from (\ref{J_N}) that
\begin{equation} \label{tr_JN}
{\rm tr}\, J_N = \int {\rm tr}\left(\vec{E}_0^{\, \dag}(\beta)U_N^{\dag}(\beta)
                         U_N^{\phantom{\dag}}(\beta)
                         \vec{E}_0^{\phantom{\dag}}(\beta)\right)\, d\beta
         = \int B(\beta) \, d\beta = 1.
\end{equation}
Thus,  the trace of the coherence matrix at any point of the fiber is
equal to~$1$,  irrespectively of the concrete structure of
inhomogeneities in the fiber.

To calculate the limit mean value of the coherence matrix,  consider
the bilinear form
\[
(J_N\vec{s}, \vec{r}\, ) =
     \int \left(\vec{s}, \vec{E}_N(\beta)\right)
          \left(\overline{\vec{r}, \vec{E}_N(\beta)}\right) d\beta,
\]
and find the limit mean value for this form for arbitrary vectors
$\vec{s}$ and $\vec{r}$ from the space of two-dimensional complex vectors
$\mathbb{C}^2$.  Since the distribution of the vector
$\vec{E}_N(\beta)$ on the three-dimensional sphere of radius $B(\beta)$
as $N\to\infty$ is  uniform (Corollary~\ref{Cor1}),  we have
\[
\langle (J_{\infty}\vec{s}, \vec{r}\, )\rangle = \int B(\beta) \, d\beta
\int_{S^3}(\vec{s}, \vec{\varsigma}\, )(\overline{\vec{r}, \vec{\varsigma}
}\, )\, d\vec{\varsigma}
  = f(\vec{s}, \vec{r}\, ),
\]
where $S^3$ is the unit three-dimensional sphere,  and $f$ is a
function of two vectors-arguments.  It is easy to show that $f$ is
invariant with respect to an arbitrary rotation specified by the
unitary matrix $V$,  i.e.,  $f(V\vec{s}, V\vec{r}\, ) = f(\vec{s}, \vec{r}\, )$.
Additionally,  this function is linear in the first argument and
antilinear in the second one.  We will show now that $f(\vec{s}, \vec{r}\, )
= \alpha(\vec{s}, \vec{r}\, )$,  where $\alpha={\rm const}$.  Fixing the first
argument of $f$ and using linearity,  we see that $f(\vec{s}, \vec{r}\, ) =
(g(\vec{s}), \vec{r}\, )$,  where $g(\cdot)$ is a linear function on a finite
dimensional vector space.  Hence,  there exists a matrix $G$ such that
${g(\vec{s}\, ) = G\vec{s}}$.  Thus,  $f(\vec{s}, \vec{r}\, ) = (G\vec{s}, \vec{r}\, )$.
Using the invariance of $f$ we have ${(GV\vec{s}, V\vec{r}\, ) =
(V^{\dag}GV\vec{s}, \vec{r}\, ) = (G\vec{s}, \vec{r}\, )}$,  i.e.,  $G = V^{\dag}G V$
for any unitary matrix $V$.  Due to the Schur lemma \cite{Group} all
eigenvalues of the matrix $G$ with this property coincide,  i.e.,  $G =
\alpha E$,  where $E$ is the unit matrix.  Thus,
$\langle (J_{\infty}\vec{s}, \vec{r}\, )\rangle = \alpha(\vec{s}, \vec{r}\, )$,  where
$\alpha$ is a constant.  To find this constant,  we calculate
the limit mean value of the trace of the coherence matrix.  Using the
formula
\[
{\rm tr}\, J_N = (J_N\vec{e}_1, \vec{e}_1) + (J_N\vec{e}_2, \vec{e}_2),
\]
where $\{\vec{e}_1, \vec{e}_2\}$ is an
orthonormal basis in $\mathbb{C}^2$,
we obtain: $\langle {\rm tr}\, J_{\infty}\rangle = 2\alpha$. But,  on the other hand,
since (\ref{tr_JN}) we deduce that ${\rm tr}\, J_{\infty} = {\rm tr}\, J_0 = 1$. Therefore,
$\alpha = \frac{1}{2}$ and
\begin{equation} \label{MJxy}
\langle (J_{\infty}\vec{s}, \vec{r}\, )\rangle = \frac{1}{2}(\vec{s}, \vec{r}\, ).
\end{equation}
It follows from (\ref{MJxy}) that $\langle J_{\infty}\rangle=\frac{1}{2}\, E$.
 From the second part of Theorem~\ref{Th2} it follows also that
$\lim\limits_{z\to\infty}\langle J(z)\rangle=\frac{1}{2}\, E$.  This is exactly
the assertion of Corollary~\ref{Cor3}.

Let us calculate now the limit mean value of the square of
polarization degree.  The square of the polarization degree at the
output of the fiber that consists of $N$ random sections is described
by the formula
\begin{equation} \label{p}
p_N^2 = 1-\frac{4\det J_N}{{\rm tr}^2 J_N}.
\end{equation}
Therefore,  taking into account (\ref{tr_JN}),  to find the limit mean
square of the degree of polarization using formula (\ref{p}),  we have
to find the limit mean value of $\det J_N$.  To this end,  we
introduce an orthonormal basis $\{\vec{e}_1, \, \vec{e}_2\}$ in
$\mathbb{C}^2$. Then
\begin{equation} \label{det_J}
\det J_N = (J_N\vec{e}_1, \vec{e}_1)(J_N\vec{e}_2, \vec{e}_2) -
|(J_N\vec{e}_2, \vec{e}_1)|^2.
\end{equation}
Clearly, to calculate $\langle\det J_N\rangle$, one should know how to find
$\langle(J_N\vec{s}, \vec{r}\,)(J_N\vec{u}, \vec{v}\,)\rangle$ for arbitrary vectors $\vec{s}$,
$\vec{r}$, $\vec{u}$ and $\vec{v}$.  We have
\[
(J_N\vec{s}, \vec{r}\, )(J_N\vec{u}, \vec{v}\, ) =
  \iint \left(\vec{s}, \vec{E}_N(\beta_1)\right)
           \left(\overline{\vec{r}, \vec{E}_N(\beta_1)}\right)
           \left(\vec{u}, \vec{E}_N(\beta_2)\right)
           \left(\overline{\vec{v}, \vec{E}_N(\beta_2)}\right)
            d\beta_1\, d\beta_2.
\]
Since the vectors $\vec{E}_N(\beta_1)$ and $\vec{E}_N(\beta_2)$ are
independent as $N\to\infty$ (Corollary~\ref{Cor2}),  and their limit
distributions are uniform (Corollary~\ref{Cor1}),  we obtain
\begin{gather*}
  \langle(J_{\infty}\vec{s}, \vec{r}\, )(J_{\infty}\vec{u}, \vec{v}\, )\rangle =
  \iint B(\beta_1) B(\beta_2) \, d\beta_1\, d\beta_2
  \int_{S^3}\int_{S^3} (\vec{s}, \vec{\varsigma}\, )
  (\overline{\vec{r}, \vec{\varsigma}}\, )
  (\vec{u}, \vec{\tau}\, )(\overline{\vec{v}, \vec{\tau}}\, )
  \, d\vec{\varsigma}\, d\vec{\tau} \\
\qquad = \left(\int
  B(\beta)\, d\beta\right)^2 \int_{S^3}(\vec{s}, \vec{\varsigma}\, )
  (\overline{\vec{r}, \vec{\varsigma}}\, )\, d\vec{\varsigma}
  \int_{S^3}(\vec{u}, \vec{\varsigma}\, )
  (\overline{\vec{v}, \vec{\varsigma}}\, )\, d\vec{\varsigma} =
  \frac{1}{4}(\vec{s}, \vec{r}\, )(\vec{u}, \vec{v}\, ).
\end{gather*}
Applying this result to relation (\ref{det_J}) and taking into
account that vectors $\vec{e}_1$ and $\vec{e}_2$ are orthonormal,  we
see that $\langle\det J_{\infty}\rangle = \frac{1}{4}$.  Thus,  it follows
from (\ref{p}) that $\langle p^2_{\infty}\rangle = 0$,  and since
$\langle p_{\infty}\rangle \leqslant\sqrt{\langle p^2_{\infty}\rangle }$,  we also have
$\langle p_{\infty}\rangle=0$.

 From the second part of Theorem \ref{Th2} it follows that
$\lim\limits_{z\to\infty}\langle p^2(z)\rangle =0$.

\subsection{Proof of Proposition \ref{Prop1}}

We wish to estimate the dependence of the diagonal
components of $J_N(\beta)$ on $N$ in the case when the
distribution of $\Theta$ is symmetric with respect to zero,  i.e.,
$\langle \Theta^{2k+1}\rangle = 0$ for all integers $k$.

Clearly,  the matrix
$J_N(\beta)=\vec{E}_N^{\vphantom{dag}}(\beta)\vec{E}_N^{\;\dag}(\beta)$ is
Hermitian.  According to our model,  the vector $\vec{E}_k$ at the
endpoint of the $k$-th fiber section results from the vector
$\vec{E}_{k-1}$ at the endpoint of $(k-1)$-st fiber section by
multiplying by the unitary matrix $M_{\beta}(l_k, \Theta_k)$ (see
(\ref{Matr_M})).  Hence,
\begin{equation} \label{MJM}
J_k(\beta) = M_{\beta}(l_k, \Theta_k)
     J_{k-1}(\beta) M_{\beta}^{\dag}(l_k, \Theta_k)\equiv
     \widehat{M}(\beta;\, l_k, \Theta_k)(J_{k-1}(\beta)).
\end{equation}
Now, consider the introduced linear operator in the space $\mathcal
H$ of Hermitian ${2{\times}2}$-matrices.  It is easy to see that the
operator family (\ref{MJM}) preserves the usual inner product
$(A, B)\equiv {\rm tr}\,(AB^\dag)$ in the space of complex matrices and,
therefore,  it is unitary.  Observe also that operators from this
family belong to the tensor product of the standard two-dimensional
representation of $SU(2)$ in $\mathbb{C}^2$ and its Hermitean
conjugate; hence,  have a unique common eigenvector
$\sigma_0=\left(\begin{array}{cc}1&0\\0&1\end{array}\right)$ with
common eigenvalue~1.  One can see also from (\ref{MJM}) that {\it if
the distribution of $\Theta$ is symmetric with respect to zero},  the
operator $\langle \widehat{M}(\beta)\rangle$ has another,  orthogonal to previous one,
eigenvector $
\sigma_1=\left(\begin{array}{rr}1&0\\0&-1\end{array}\right)$.  Then
the subspace of the diagonal matrices is invariant with respect to the
operator $\langle \widehat{M}(\beta)\rangle$.  Hence,  we obtain the relation
\begin{equation} \label{DiagJ}
   \mbox{diag}\,(\langle J_N(\beta)\rangle)=
   \frac{1}{2}\, \left(\mbox{diag}\,(J_0(\beta)), \, \sigma_0\right)\, \sigma_0+
   \frac{1}{2}\, \eta_1^N
\left(\mbox{diag}\,(J_0(\beta)), \, \sigma_1\right)\, \sigma_1,
\end{equation}
where $\eta_1$ is the eigenvalue of the operator $\langle\widehat{M}(\beta)\rangle$
corresponding to the eigenvector $\sigma_1$.  Clearly,
\[
\eta_1=\frac{1}{2}\left(\langle \widehat{M}(\beta)\rangle\sigma_1, \sigma_1 \right)
=\frac{1}{2}\left\langle {\rm tr}\, M_\beta(l, \Theta)\sigma_1
M_\beta^\dag(l, \Theta)\sigma_1\right\rangle=1-2\left\langle m_{\beta 3}^2(l, \Theta)\right\rangle.
\]
It follows now from (\ref{DiagJ}) that
\begin{gather}
\langle J_{11}(\beta, N)\rangle = \frac{{\rm tr}\, J_0}{2}
     \left(1+\frac{J_{11}^0-J_{22}^0}{{\rm tr}\, J_0}\, \eta_1^N\right), \nonumber\\
\langle J_{22}(\beta, N)\rangle = \frac{{\rm tr}\, J_0}{2}
     \left(1-\frac{J_{11}^0-J_{22}^0}{{\rm tr}\, J_0}\, \eta_1^N\right) .\label{PN}
\end{gather}
If $\eta_1>0$,  this relation can be re-written in the form similar to
(\ref{J_diag}):
\begin{gather}
\langle J_{11}(\beta, z)\rangle = \frac{{\rm tr}\, J_0}{2}
     \left(1+\frac{J_{11}^0-J_{22}^0}{{\rm tr}\, J_0}\, {\rm e}^{-2hz}\right), \nonumber\\
\langle J_{22}(\beta, z)\rangle = \frac{{\rm tr}\, J_0}{2}
     \left(1-\frac{J_{11}^0-J_{22}^0}{{\rm tr}\, J_0}\, {\rm e}^{-2hz}\right) ,\label{PZ}
\end{gather}
where $z=N\langle l\rangle$ and $h=-\frac{1}{2\langle l\rangle}\ln\eta_1$.

\subsection{Proof of Proposition \ref{Prop2}}

As in the proof of Theorem \ref{Th2},  to calculate
the value of the $h$-parameter,  we use the
Laplace transformation.  Let $N(z)$ be defined in
(\ref{N(z)}),  $z_k=\sum\limits_{j=1}^{k}l_j$ and $\Phi_l(z)$ be
defined in (\ref{Phi_l}).  We deduce from (\ref{h_class}),  due to
independence of random values $l_k,  \Theta_k$,  that
\begin{gather*}
\hspace*{-9mm} h(\beta)=\lim\limits_{z\to\infty}\frac{1}{z}\left\langle \sum_{k=0}^{N(z)-1}
\!\int_{z_k}^{z_{k+1}}\!\!\int_{z_k}^{z_{k+1}}\Theta_{k+1}^2
{\rm e}^{{\rm i}\beta(t-s)}\, dt\, ds+\int_{z_{N(z)}}^z\!\!\int_{z_{N(z)}}^z
\Theta_{N(z)+1}^2{\rm e}^{{\rm i}\beta(t-s)}\, dt\, ds\right\rangle\\
=
\frac{4\langle \Theta^2\rangle }{\beta^2}\lim\limits_{z\to\infty}\frac{1}{z}
\sum\limits_{k=0}^{\infty}  \int_{z_k\leqslant z}
\prod\limits_{j=1}^{k}\rho_l(l_j)\, dl_j\, \Phi_l(z-z_k)
\left(\sum\limits_{j=1}^{k}\sin^2\frac{l_j\beta}{2}+\sin^2
\frac{(z-z_k)\beta}{2}\right).
\end{gather*}
Using now the convolution theorem for the Laplace transformation,  we
can express the last expression as
\begin{gather*}
h(\beta)=\frac{4\langle \Theta^2\rangle}{\beta^2}\lim\limits_{z\to\infty}
\frac{1}{2\pi{\rm i}\, z} \int_{a-{\rm i}\infty}^{a+{\rm i}\infty}\left(
\frac{\widehat{\Psi}_l(\lambda)}{1-\hat{\rho}_l(\lambda)}+
\frac{\widehat{\Phi}_l(\lambda)\hat{\psi}_l(\lambda)}{(1-\hat{\rho}_l
(\lambda))^2}
\right) d\lambda\\
\phantom{h(\beta)} =
\frac{4\langle \Theta^2\rangle }{\beta^2}\lim\limits_{z\to\infty}\frac{1}{2\pi{\rm
i}} \int_{a-{\rm i}\infty}^{a-{\rm i}\infty}\lambda\left(
\frac{\widehat{\Psi}_l(\lambda)}{1-\hat{\rho}_l(\lambda)}+
\frac{\widehat{\Phi}_l(\lambda)\hat{\psi}_l(\lambda)}{(1-\hat{\rho}_l
(\lambda))^2}
\right){\rm e}^{\lambda z}\, d\lambda\\
\phantom{h(\beta)} =
\frac{4\langle \Theta^2\rangle }{\beta^2}\,{\mathop {\rm Res}\limits_{\lambda=0}}
\left(
\frac{\lambda\widehat{\Psi}_l(\lambda)}{1-\hat{\rho}_l(\lambda)}+
\frac{\lambda\widehat{\Phi}_l(\lambda)\hat{\psi}_l(\lambda)}{(1-\hat{
\rho}_l(\lambda))^2}
\right),
\end{gather*}
where $\hat{\rho}_l(\lambda)$ is defined in (\ref{hat_rho}) and
\begin{gather*}
\widehat{\Phi}_l(\lambda)=\int_{0}^{\infty}\Phi_l(z){\rm
e}^{-\lambda z}\, dz=
\frac{1}{\lambda}(1-\hat{\rho}_l(\lambda)), \\
\widehat{\Psi}_l(\lambda)=\int_{0}^{\infty}\Phi_l(z)\sin^2
\frac{z\beta}{2}\, {\rm e}^{-\lambda z}\, dz, \\
\hat{\psi}_l(\lambda)=\int_{0}^{\infty}\rho_l(z)\sin^2\frac
{z\beta}{2}\,
{\rm e}^{-\lambda z}\, dz.
\end{gather*}
We see that $\widehat{\Psi}_l(\lambda)$,   $\hat{\psi}_l(\lambda)$
and $\frac{\lambda}{1-\hat{\rho}_l(\lambda)}$ have  finite
limits as $\lambda\to 0$. Hence,
\[
{\mathop{\rm Res}\limits_{\lambda=0}}\left(
\frac{\lambda\widehat{\Psi}_l(\lambda)}{1-\hat{\rho}_l(\lambda)}+
\frac{\lambda\widehat{\Phi}_l(\lambda)\hat{\psi}_l(\lambda)}{(1-\hat{
\rho}_l(\lambda))^2}
\right)=\frac{\hat{\psi}_l(0)}{\displaystyle -\frac{d\hat{\rho}_l(\lambda)}{d\lambda}
\Bigr|_{\lambda=0}}=
\frac{1}{\langle l\rangle} \left\langle \sin^2\frac{l\beta}{2}\right\rangle.
\]
Thus,  we obtain relation (\ref{hClassEx}).

\subsection{Proof of Proposition \ref{Prop3}}

In this section we consider an asymptotic expansion
of the mean square of the polarization degree at large $N$.  We can
only find the principal term of this asymptotic.

Note that simultaneously with the calculation of the asymptotic
expansion of the mean square of the polarization degree we obtain
another proof of the relation $\langle p^2_{\infty}\rangle = 0$.

Our problem reduces to the asymptotic estimate of the mean value
of the determinant of the coherence matrix.  To do this,  it suffices
to calculate the mean value the tensor product $J_N\otimes J_N$,
because the determinant of $2{\times}2$-matrix is a {\it linear}
function of the elements of its tensor square.  Further,  since $J_N$
is an integral over the parameter $\beta$ of matrices $J_N(\beta)$,
the mentioned tensor product is the integral over the pair $(\beta_1,
\beta_2)$ of the tensor product $J_N(\beta_1)\otimes J_N(\beta_2)$.
It follows from (\ref{MJM}) that there exists a linear
operator $\widehat{M}(\beta_1, \beta_2;\, l, \Theta)$ in
$\mathcal{H}^2\equiv\mathcal{H}\otimes\mathcal{H}$ such that
\begin{equation}\label{M2}
   J_N(\beta_1)\otimes J_N(\beta_2)=\widehat{M}(\beta_1, \beta_2;\, l_N, \Theta_N)
                                 \left(J_{N-1}(\beta_1)\otimes
J_{N-1}(\beta_2)\right).
\end{equation}
Note that the operator family $\widehat{M}(\beta_1, \beta_2;\, l, \Theta)$ is
unitary because it preserves the inner product ${(A_1\otimes
B_1, A_2\otimes B_2)\equiv (A_1, A_2)(B_1, B_2)}$ in $\mathcal{H}^2$.

Hereafter we denote by $\left\{\sigma_j, \
j\in\{0, 1, 2, 3\}\right\}$ the usual Pauli basis in the space
$\mathcal{H}$ of Hermitian $2\times 2$-matrices:
\[
\sigma_0=\left(\begin{array}{rr}1&0\\0&1\end{array}\right), \qquad
\sigma_1=\left(\begin{array}{rr}1&0\\0&-1\end{array}\right), \qquad
\sigma_2=\left(\begin{array}{rr}0&1\\1&0\end{array}\right), \qquad
\sigma_3=\left(\begin{array}{rr}0&i\\-i&0\end{array}\right).
\]
We also denote by $s_j=\frac{\sigma_j}{\sqrt{2}}$ the corresponding
orthonormal basis.  Remember that the inner product in $\mathcal{H}$
is defined as $(A, B)={\rm tr}\, AB$.  For the set $\{x_\alpha\}$ of vectors
in a given vector space we denote by
$\mathcal{L}\left(\{x_\alpha\}\right)$ its linear envelope.

Let us now list some facts from the representation theory of the
group $SU(2)$.

F1) Let $\pi$ be a representation of the group $SU(2)$ in the space $\mathcal{H}$ defined by
the relation $\pi(g)A=gAg^\dag$, $A\in\mathcal{H}$.  Clearly,  this
representation is reducible and its decomposition into irreducible
components is of the form $\mathcal{H}=H_0\oplus H_3$,  where
$H_0=\mathcal{L}({\sigma_0})$ and
$H_3=\mathcal{L}\left(\{\sigma_1, \sigma_2, \sigma_3\}\right)$.

F2) Let $\pi^{\otimes 2}$ be a representation of
$SU(2)\times SU(2)$ in $\mathcal{H}^2$ such that
\[
\pi^{\otimes 2}(g_1, g_2)(A\otimes B)=\pi(g_1)A\otimes \pi(g_2)B.
\]
Then from the decomposition above it follows that
\[
\mathcal{H}^2=H_0\otimes
H_0\oplus H_0\otimes H_3\oplus H_3\otimes H_0\oplus H_3\otimes
H_3\equiv H^2_{00}\oplus H^2_{03}\oplus H^2_{30}\oplus H^2_{33}
\]
and this is the decomposition of $\pi^{\otimes 2}$ into irreducible
components.

F3) Denote by $\pi^2$ the representation of $SU(2)$ in
$\mathcal{H}^2$,  obtained by restricting $\pi^{\otimes 2}$ onto the
image of the diagonal embedding of $SU(2)$ into $SU(2)\times SU(2)$.
Then the components $H^2_{00}$,
$H^2_{03}$ and $H^2_{30}$ are still irreducible,  but the component $
H^2_{33}$ can be further decomposed.  Indeed,  consider the natural
action of the group $\mathbb{Z}_2=\{\bar{0}, \,  \bar{1}\}$ on
$\mathcal{H}^2$ by twist:  ${\bar{1}(A\otimes B)=B\otimes A}$.
Then $H^2_{33}=H^2_{33s}\oplus H^2_{33a}$ is the direct sum of the
spaces of symmetric and skew-symmetric tensors under this action.  The
space $H^2_{33s}$ can be further decomposed into the direct sum
\[
H^2_{33s}=\mathcal{L}\left(\left\{\sum\limits_{i=1}^3
\sigma_i\otimes\sigma_i\right\}\right)
\oplus\mathcal{L}\left(\left\{\sum\limits_{i=1}^3
\sigma_i\otimes\sigma_i\right\}\right)^\bot
\equiv H^2_1\oplus H^2_5.
\]
These are all irreducible components of $\pi^2$.

Thus,  the complete decomposition of $\pi^2$ in $\mathcal{H}^2$
consists of {\it two} one-dimensional components,  spanned by vectors
$\varepsilon_0=s_0\otimes s_0$ and
$\varepsilon_1=\sum\limits_{i=1}^3s_i\otimes s_i$,  {\it three}
three-dimensional
components in the spaces $H^2_{03}$, $H^2_{30}$ and $H^2_{33a}$ and {\it one}
five-dimensional component in $ H^2_5$.

We will also need some relations between vectors and operators
in $\mathcal{H}^2$.

R1) Let $T=\varepsilon_0+\varepsilon_1$. Then for
$A, B\in\mathcal{H}$,  one has $(A\otimes B,  T)={\rm tr}\, AB$.

R2) $\Delta=\frac{1}{2}\, (\varepsilon_0-\varepsilon_1)$.  Then for
$A\in\mathcal{H}$,  one has $(A\otimes A, \Delta)=\det A$.

R3) Let $g: \mathbb{R}\longrightarrow SU(2)$ be a smooth
function, and $g_2: \mathbb{R}^2\longrightarrow SU(2)\times SU(2)$ be
such that $g_2(x, y)=(g(x), g(y))$.  Let further $D$ be a differential
operator in $C^\infty(\mathbb{R}^2)$ of the form
$D=\frac{\partial}{\partial x}-\frac{\partial}{\partial y}$.  (Here
we denote by $C^\infty(\mathbb{R}^2)$ the space of smooth functions
with values in an arbitrary finite dimensional vector space.)  The
operator $D_0: C^\infty(\mathbb{R}^2)\longrightarrow
C^\infty(\mathbb{R})$ is then $\mathop{D|}_{y=x}$, and similarly
$D_0^2=\mathop{D^2|}_{y=x}$.  The following relations can be proven by
direct, but long, calculations:
\begin{gather}
D_0^2\left(\pi^{\otimes 2}(g_2(x, y))\varepsilon_1, \, \varepsilon_1\right)=
-4\,{\rm tr}\,\frac{dg^\dag}{dx}\, \frac{dg}{dx}, \label{Sder}\\
D_0\pi^{\otimes 2}(g_2(x, y))\varepsilon_1\in H^2_{33a}. \label{Fder}
\end{gather}

Denote by $\widehat {L}(\beta_1, \beta_2)$ the mean operator
$\langle \, \widehat{M}(\beta_1, \beta_2;\, l, \Theta)\rangle$ and by $F_N(\beta_1, \beta_2)$
the mean value of $J_N(\beta_1)\otimes J_N(\beta_2)$.  Then it follows
from (\ref{M2}) that
\begin{equation}\label{MidM2}
   F_N(\beta_1, \beta_2)=\widehat{L}^N(\beta_1, \beta_2) F_0(\beta_1, \beta_2),
\end{equation}
and we can use,  as in the previous section,  the spectral decomposition
of the operator $\widehat{L}(\beta_1, \beta_2)$ to calculate
$F_N(\beta_1, \beta_2)$.

Note that we deal here with operators in the $16$-dimensional space
$\mathcal{H}^2$ and there are no reasonable reasons for existence of
an analytical solution of spectral problem for operator
$\widehat{L}(\beta_1, \beta_2)$.  This is the main obstacle for obtaining
complete asymptotic decomposition of the polarization degree.  We will
see,  nevertheless,  that it is possible to obtain some analytical
expressions for two major eigenvalues and corresponding eigenvectors
of this operator which suffices to construct the leading term of the
asymptotic of the polarization degree.

Using definitions above,  we deduce:
\begin{gather}
\langle \det J_N\rangle=
\iint\left(F_N(\beta_1, \beta_2), \Delta\right)\, d\beta_1d\beta_2\nonumber\\
\phantom{\langle \det J_N\rangle} =\iint\left(\widehat{L}^N(\beta_1, \beta_2) F_0(\beta_1, \beta_2), \Delta\right)
    \, d\beta_1d\beta_2. \label{MDetJN}
\end{gather}
The integrand in (\ref{MDetJN}) can be written in the form
\begin{gather}
\left(\widehat {L}^N(\beta_1, \beta_2) F_0(\beta_1, \beta_2), \Delta\right)=\frac{1}{2}
\left(F_0(\beta_1, \beta_2), \left(\widehat{L}^\dag(\beta_1, \beta_2)\right)^N
      (\varepsilon_0-\varepsilon_1)\right)\nonumber\\
\qquad =
\frac{1}{2}\left(F_0(\beta_1, \beta_2), \varepsilon_0\right)-\frac{1}{2}
  \left(F_0(\beta_1, \beta_2), \left(\widehat{L}^\dag(\beta_1, \beta_2)\right)^N\varepsilon_1
\right). \label{LFD}
\end{gather}
As follows from the definition of vector $\varepsilon_0$,
\begin{equation}\label{LFD1}
\left(F_0(\beta_1, \beta_2), \varepsilon_0\right)=
\frac{1}{2}\,{\rm tr}\, J_0(\beta_1)\,{\rm tr}\,
J_0(\beta_2)=\frac{1}{2}B(\beta_1)B(\beta_2).
\end{equation}
Let
$\{\varepsilon_i^\dag(\beta_1, \beta_2), \, \eta_i(\beta_1, \beta_2)\}_{i=0}^{15}$
be the set of normalized eigenvectors and corresponding eigenvalues of
$\widehat{L}^\dag(\beta_1, \beta_2)$.  Then the dual basis is the set
$\{\varepsilon_i(\beta_1, \beta_2)\}_{i=0}^{15}$ of
eigenvectors of $\widehat{L}(\beta_1, \beta_2)$.  Moreover,  it follows from
previous discussions that the vector
$\varepsilon_0^\dag(\beta_1, \beta_2)=\varepsilon_0(\beta_1, \beta_2)=\varepsilon_0$
is orthogonal to all other eigenvectors $\varepsilon_i$,
$\eta_0(\beta_1, \beta_2)=1$,
$\varepsilon_1^\dag(\beta, \beta)=\varepsilon_1(\beta, \beta)=
\frac{1}{\sqrt{3}}\, \varepsilon_1$ and $\eta_1(\beta, \beta)=1$.
Further,  it follows from Theorem~\ref{Th2} that
$|\eta_i(\beta_1, \beta_2)|\leqslant \nu<1$ for $i=2, \dots, 15$.
Expanding $\varepsilon_1$ in terms of the eigenvectors
$\varepsilon_i^\dag$ we obtain
\begin{gather*}
\left(\widehat{L}^\dag(\beta_1, \beta_2)\right)^N\varepsilon_1 =
\sum\limits_{i=0}^{15}\left(\varepsilon_1,
\varepsilon_i(\beta_1, \beta_2)\right)
\left(\widehat{L}^\dag(\beta_1, \beta_2)\right)^N\varepsilon_i^\dag(\beta_1, \beta_2)\\
\qquad =
\sum\limits_{i=1}^{15}\eta_i^N(\beta_1, \beta_2)
\left(\varepsilon_1, \varepsilon_i(\beta_1, \beta_2)\right)\,
\varepsilon_i^\dag(\beta_1, \beta_2).
\end{gather*}
It follows from the properties of eigenvalues $\eta_i$ that the
relation
\begin{equation} \label{Le1}
    \left(\widehat{L}^\dag(\beta_1, \beta_2)\right)^N\varepsilon_1 = {\rm
    e}^{-Nh(\beta_1, \beta_2)}
    \left(\varepsilon_1, \varepsilon_1(\beta_1, \beta_2)\right)\,
    \varepsilon_1^\dag(\beta_1, \beta_2) + O\left({\rm e}^{-\alpha N}\right),
\end{equation}
holds for $\alpha=-\ln\nu>0$ and
$h(\beta_1, \beta_2)=-\ln{\eta_1(\beta_1, \beta_2)}$.

Substituting (\ref{Le1}) into (\ref{LFD}) and using (\ref{LFD1}) we
obtain
\begin{gather}
\left(\widehat{L}^N(\beta_1, \beta_2) F_0(\beta_1, \beta_2), \Delta\right)
=
\frac{1}{4}\, B(\beta_1)B(\beta_2)\nonumber\\
\qquad {}-\frac{1}{2}\, {\rm e}^{-Nh(\beta_1, \beta_2)}
\left(\varepsilon_1, \varepsilon_1(\beta_1, \beta_2)\right)
\left(F_0(\beta_1, \beta_2), \varepsilon_1^\dag(\beta_1, \beta_2)\right)
+ O\left({\rm e}^{-\alpha N}\right).\label{LFD+}
\end{gather}
Now by substituting (\ref{LFD+}) into (\ref{MDetJN}),  changing
variables
\[
\beta=\frac{1}{2}(\beta_1+\beta_2) , \qquad
\delta=\beta_1-\beta_2,
\]
and integrating over $\delta$ by the saddle-point method,  we obtain
\begin{equation}\label{int}
\langle\det J_N\rangle = \frac{1}{4}\left(\int B(\beta)\, d\beta\right)^2\!\! -
\frac{1}{6}\int\!\left(F_0(\beta, \beta), \varepsilon_1\right)
\sqrt{\frac{2\pi}{N f(\beta)}}\:d\beta +
O\left(\frac{1}{N^{3/2}}\right),
\end{equation}
where $f(\beta)=D^2_0 h(\beta, \beta)$. Here we use the relation
\begin{equation}\label{MuDer}
D_0h(\beta,  \beta)= -D_0\eta_1(\beta,  \beta)= -\frac{1}{3}\,
D_0\left(\widehat{L}^\dag(\beta, \beta)\varepsilon_1, \, \varepsilon_1\right)=0,
\end{equation}
which follows from the standard perturbation theory \cite{Kato} and the
fact that
\[
\left(\widehat{L}^\dag(\beta_1, \beta_2)\varepsilon_1, \, \varepsilon_1\right)=
\left(\widehat{L}^\dag(\beta_2, \beta_1)\varepsilon_1, \, \varepsilon_1\right).
\]
(To prove the last relation it suffices to note that,  by definition,
\begin{gather*}
\left(\widehat {L}^\dag(\beta_1, \beta_2)(A\otimes A), \, A\otimes A\right)\\
\qquad=
\left\langle \left(M_{\beta_1}(l, \Theta)AM^\dag_{\beta_1}(l, \Theta)\otimes
M_{\beta_2}(l, \Theta)AM^\dag_{\beta_2}(l, \Theta), \, A\otimes
A\right)\right\rangle \\
\qquad = \left\langle {\rm tr}\,\left(M_{\beta_1}(l, \Theta)AM^\dag_{\beta_1}(l, \Theta)A\right)
{\rm tr}\,\left(M_{\beta_2}(l, \Theta)A\widehat{M}^\dag_{\beta_2}(l, \Theta)A\right)\right\rangle
\end{gather*}
for any Hermitian $2\times 2$-matrix $A$.)  Taking into account the
relations $\int B(\beta)\, d\beta={\rm tr}\, J_N$ and
$\varepsilon_1=\varepsilon_0-2\Delta$,  we can express (\ref{int}) in
the form
\[
\frac{1}{4}\,{\rm tr}^2 J_N-\langle \det J_N\rangle =\frac{1}{3}
\int\left(\frac{1}{4}\, {\rm tr}^2J_0(\beta)-\det J_0(\beta)\right)
\sqrt{\frac{2\pi}{N f(\beta)}}\:d\beta +
O\left(\frac{1}{N^{3/2}}\right).
\]
Using now the formula (\ref{p}) and a normalized spectral function
\[
\widetilde{B}(\beta)=\frac{{\rm tr}\, J_0(\beta)}{{\rm tr}\, J_N}=
 \frac{B(\beta)}{\int B(\beta)\, d\beta} ,
\]
we obtain expression (\ref{asymptotics}) for the polarization degree.

Our last problem is to calculate the value $f(\beta)$ in terms
of parameters of $\Theta(z)$.  To do this,  observe that
\[
D_0^2\eta_1(\beta, \beta)=D_0^2 {\rm e}^{-h(\beta, \beta)}=-D_0^2
h(\beta, \beta),
\]
because $D_0 h(\beta, \beta)=0$.  Let now apply $D_0^2$ to the
relation
\begin{equation}\label{Eigen}
   \widehat{L}^\dag(\beta_1, \beta_2)\varepsilon_1(\beta_1, \beta_2)
=\eta_1(\beta_1, \beta_2)\varepsilon_1(\beta_1, \beta_2)
\end{equation}
and then scalar it by $\varepsilon_1$.  We obtain:
\begin{gather}
f(\beta)=D_0^2 h(\beta, \beta)\nonumber\\
\qquad =
-\frac{1}{3}\left(D_0^2\widehat{L}^\dag(\beta, \beta)\varepsilon_1, \, \varepsilon_1\right)-
\frac{2}{\sqrt{3}}
\left(D_0\widehat{L}^\dag(\beta, \beta)D_0\varepsilon_1(\beta, \beta), \, \varepsilon_1\right).
\label{FinDer}
\end{gather}
The first term in the right hand side of (\ref{FinDer}) is calculated by
using relation (\ref{Sder}),
\begin{gather}
\left(D_0^2\widehat{L}^\dag(\beta, \beta)\varepsilon_1, \, \varepsilon_1\right)=
\left\langle D_0^2\left(\pi^{\otimes 2}(M_\beta(l, \Theta), M_\beta(l, \Theta))
\varepsilon_1, \, \varepsilon_1\right)\right\rangle\nonumber\\
\qquad =-4\left\langle {\rm tr}\, \frac{M_\beta^\dag(l, \Theta)}{\partial\beta}\,
\frac{M_\beta(l, \Theta)}{\partial\beta}\right\rangle.\label{FTerm}
\end{gather}
To calculate the second term,  we need to find the vector
$D_0\varepsilon_1(\beta, \beta)$.  Applying $D_0$ to
equation (\ref{Eigen}) and taking into account relation
(\ref{MuDer}),  we obtain for this vector the equation
\begin{equation}\label{EDer}
   \left(E-\widehat{L}^\dag(\beta, \beta)\right)D_0\varepsilon_1(\beta, \beta)=
   \frac{1}{\sqrt{3}}\:D_0\widehat{L}^\dag(\beta, \beta)\varepsilon_1.
\end{equation}
Due to relation (\ref{Fder}),  the right hand side of equation
(\ref{EDer}) belongs to the space $H^2_{33a}$,  which is invariant,  as
follows from~F3),  under the operator in the left hand side.  The
space $H^2_{33a}$ is,  by definition,  orthogonal to the spaces
$H^2_{00}$ and $H^2_1$ (see~F2) and F3)).  The direct sum
of these spaces is exactly the kernel of the operator
$E-\widehat{L}^\dag(\beta, \beta)$,  hence,  the equation (\ref{EDer}) has a
unique solution which is orthogonal to $H^2_{00}\oplus H^2_1$,  and
this solution belongs to $H^2_{33a}$.  Since vector
$\varepsilon_1(\beta_1, \beta_2)$ is normalized and orthogonal to
$\varepsilon_0$,  this solution is the one we need.  So,  we can rewrite
equation (\ref{EDer}) as an equation in the three-dimensional
space $H^2_{33a}$.  It is easy to see that the orthonormalized basis
in $H^2_{33a}$ can be chosen in the form
\[
\phi_1=\frac{s_1\otimes s_2-s_2\otimes s_1}{\sqrt{2}}, \qquad
\phi_2=\frac{s_2\otimes s_3-s_3\otimes s_2}{\sqrt{2}}, \qquad
\phi_3=\frac{s_3\otimes s_1-s_1\otimes s_3}{\sqrt{2}}.
\]
Introduce the $3\times 3$-matrix
$S_{ij}=\left(\left(E-\widehat{L}^\dag(\beta, \beta)\right)\phi_j, \phi_i\right)$,
$i, j\in\{1, 2, 3\}$,  and vectors
$v_i=\left(D_0\varepsilon_1(\beta, \beta), \phi_i\right)$ and
$u_i=\left(D_0\widehat{L}^\dag(\beta, \beta)\varepsilon_1, \phi_i\right)$,
$i\in\{1, 2, 3\}$.  Then equation (\ref{EDer}) takes the form
\begin{equation}\label{V}
   S v=\frac{1}{\sqrt{3}}\, u.
\end{equation}
By direct calculation we see that
\[
S=2\left(\begin{array}{rrr}
         m_{11}&  m_{13}&  m_{01}\\
        m_{13}&   m_{33}&  m_{03}\\
       - m_{01}& - m_{03}& 1- m_{00}
         \end{array}\right), \qquad
          u=4\sqrt{2}\left(\begin{array}{r}
             d_{30}\\ -d_{10}\\ -d_{31}
                  \end{array}\right).
\]
Here
\begin{gather*}
m_{kj}=\langle m_{\beta k}(l, \Theta)m_{\beta j}(l, \Theta)\rangle,\\
d_{kj}=\left\langle \frac{\partial m_{\beta k}(l, \Theta)}{\partial
\beta}m_{\beta j}(l, \Theta) -m_{\beta k}(l, \Theta)\frac{\partial
m_{\beta j}(l, \Theta)}{\partial \beta}\right\rangle, \\
m_{\beta k}(l, \Theta)=\frac{{\rm i}^k}{2}\,{\rm tr}\,{M_\beta(l, \Theta)\sigma_k}, \qquad
k, j\in\{0, 1, 3\}.
\end{gather*}
We can now express the function $f(\beta)$ in terms of components of
vector $v$ and elements of matrix $M_\beta(l, \Theta)$ and its
derivatives with respect to $\beta$.  Using (\ref{FinDer}),
(\ref{FTerm}) we finally obtain
\[
f(\beta)=\frac{8}{3}\sum\limits_{k\in\{0, 1, 3\}}
\left\langle \left(\frac{\partial m_{\beta j}(l, \Theta)}{\partial
\beta}\right)^2\right\rangle- 8 \sqrt{\frac{2}{3}}\, (-v_1 d_{30}+v_2 d_{10}-v_3
d_{31})
\]
This relation is essentially simplified when the regular twist is
absent: $u$ and many elements of $S$ vanish and the solution of
equation (\ref{V}) attains a simple explicit form:
\[
v=\left(\begin{array}{c}
     0\vspace{1mm}\\
     \displaystyle -2\sqrt{\frac{2}{3}}\, \frac{d_{10}}{m_{33}}\vspace{1mm}\\
     0
   \end{array}
   \right).
\]
So,  in this case
\begin{gather*}
f(\beta)=\frac{8}{3}\left\langle \frac{l^2\beta^2}{4(\beta^2+4\Theta^2)}+
\frac{\displaystyle 4\Theta^2\sin^2\frac{l}{2}\sqrt{\beta^2+4\Theta^2}}
{(\beta^2+4\Theta^2)^2}\right\rangle\\
\phantom{f(\beta)=}+\frac{4}{3}
\frac{\displaystyle \left\langle \frac{l\beta^2}{\beta^2+4\Theta^2}+\frac{4\Theta^2\sin
l\sqrt{\beta^2+4\Theta^2}} {(\beta^2+4\Theta^2)^{3/2}}\right\rangle^2}
{\displaystyle \left\langle \frac{\Theta^2}{\beta^2+4\Theta^2}\sin^2\frac{l}{2}\sqrt{\beta^2
+4\Theta^2}\right\rangle}.
\end{gather*}
This completes the proof of Proposition~\ref{Prop3}.

\subsection*{Acknowledgements}

We are thankful to Vl~V~Kocharovsky for helpful discussions and
D~A~Leites for encoura\-gement and help.  The first author is thankful
also for RFBR grants $N^{\underline{o}}$~00-15-96732 and
$N^{\underline{o}}$~00-02-17344 for partial financial support.

\label{malykin-lastpage}

\end{document}